\documentclass[twocolumn,appendixfloats]{aastex62}
\pdfoutput=1 %for arXiv submission
\usepackage{amsmath,amstext}
\usepackage[T1]{fontenc}
\usepackage{apjfonts} 
\usepackage[figure,figure*]{hypcap}
\usepackage{float}
\usepackage{hyperref}
\hypersetup{
    colorlinks=true,
    linkcolor=cyan,
    citecolor=cyan,
    filecolor=cyan,      
    urlcolor=cyan,
}

\shorttitle{AGN All the Way Down?}
\shortauthors{Dickey et al.}

\begin{document}

\newcommand{\mstel}{\mathrm{M_*}}
\newcommand{\msun}{\mathrm{M_\odot}}
\newcommand{\todo}[1]{{\bf \textcolor{red}{ #1}}}
\newcommand{\ha}{H$\mathrm{\alpha}$}
\newcommand{\hb}{H$\beta \,$}
\newcommand{\dnfour}{$\mathrm{D_n4000}$}
\newcommand{\othreehb}{[OIII]/H$\beta \,$}
\newcommand{\dbpt}{$\mathrm{d_{BPT}}$}
\newcommand{\ntwo}{$\mathrm{N}\, \textsc{ii}$}.pdf
\newcommand{\othree}{$\mathrm{O}\, \textsc{iii}$}
\newcommand{\htwo}{$\mathrm{H}\, \textsc{ii}$}
\newcommand{\stwo}{$\mathrm{S}\, \textsc{ii}$}

\title{AGN All the Way Down? \\AGN-like Line Ratios are Common In the Lowest-Mass Isolated Quiescent Galaxies}

\author{Claire Mackay Dickey}
\affiliation{Department of Astronomy, Yale University, New Haven, CT 06520, USA}

\author{Marla Geha}
\affiliation{Department of Astronomy, Yale University, New Haven, CT 06520, USA}

\author{Andrew Wetzel}
\altaffiliation{Caltech-Carnegie Fellow}
\affiliation{Department of Physics, University of California, Davis, CA 95616, USA}
\affiliation{TAPIR, California Institute of Technology, Pasadena, CA 91125, USA}
\affiliation{The Observatories of the Carnegie Institution for Science, 813 Santa Barbara St, Pasadena, CA 91101, USA}

\author{Kareem El-Badry}
\affiliation{Department of Astronomy and Theoretical Astrophysics Center, University of California Berkeley, Berkeley, CA 94720, USA}

\email{claire.dickey@yale.edu}

%\received{1 February 2019}
%\submitjournal{ApJ}

\begin{abstract}

    We investigate the lowest-mass quiescent galaxies known to exist in isolated environments ($\mathrm{M^* = 10^{9.0-9.5} \ M_\odot}$; 1.5\,Mpc from a more massive galaxy).  This population may represent the lowest stellar mass galaxies in which internal feedback quenches galaxy-wide star formation.  We present Keck/ESI long-slit spectroscopy for 27 isolated galaxies in this regime:  20 quiescent galaxies and 7 star-forming galaxies.  We measure emission line strengths as a function of radius and place galaxies on the Baldwin Phillips Terlevich (BPT) diagram.  Remarkably, 16 of 20 quiescent galaxies in our sample host central AGN-like line ratios.  Only 5 of these quiescent galaxies were identified as AGN-like in SDSS due to lower spatial resolution and signal-to-noise.  We find that many of the quiescent galaxies in our sample have spatially-extended emission across the non-SF regions of BPT-space.  When considering only the central 1\arcsec, we identify a tight relationship between distance from the BPT star-forming sequence and host galaxy stellar age as traced by \dnfour, such that older stellar ages are associated with larger distances from the star-forming locus.  Our results suggest that the presence of hard ionizing radiation (AGN-like line ratios) is intrinsically tied to the quenching of what may be the lowest-mass self-quenched galaxies.

\end{abstract}


\section{Introduction}

    Understanding the processes that regulate star formation and quench galaxies remains a major goal for studies of galaxy evolution.  Since the advent of the hierarchical CDM paradigm, a feedback mechanism or mechanisms have often been invoked to explain the discrepancies between the observed stellar mass function and the halo mass function.  \citep{white1978,dekel1986,white1991}. Feedback mechanisms fall into two categories: processes that arise in high-density environments and those that are internal to galaxies \citep{peng2010}.  A diverse set of external phenomena (ram-pressure stripping, tidal forces, major and minor mergers, etc.) are able to remove or heat gas in group and cluster environments and are particularly effective in low-mass galaxies \citep[e.g.,][]{pasquali2010,smith2012}.  Within individual galaxies, active galactic nuclei (AGN) appear necessary to produce the observed population of massive quenched galaxies \citep[e.g.,][]{croton2006,somerville2008,choi2015,su2018}, heating gas via high-velocity winds and radio jets.  The efficiency of internal versus external quenching mechanisms is a strong function of galaxy mass; with high-mass galaxies dominated by processes internal to themselves and low-mass galaxies exclusively quenched by environmental effects \citep{somerville2015}. 
    
    While environmental processes can effectively quench galaxies across almost all mass regimes \citep[the most massive galaxies being an exception;][]{peng2010, wetzel2013}, the efficiency of self-quenching is thought to be tightly correlated with mass. \citet{geha2012} found that there is a stellar mass threshold below which isolated galaxies at $z < 0.055$ in the Sloan Digital Sky Survey \citep[SDSS;][]{aihara2011} cannot quench themselves.  They found that all of the several thousand galaxies with $\mstel = 10^{7.0-9.0} \ \msun$ identified in SDSS as isolated (defined as 1.5\,Mpc from a more massive companion), were star-forming.  Whatever process(es) act within low-mass galaxies to disrupt star formation either become inefficient or cannot occur below $\mstel = 10^{9.0} \ \msun$.  This quenching threshold is also observed to be consistent across redshift \citep{papovich2018}.  Supernovae, stellar winds, and AGN are all possible sources for driving feedback in galaxies just above this quenching threshold, but their relative impact on the evolution of low-mass galaxies remains uncertain.  While it is widely agreed that AGN are both ubiquitous and influential in massive ($\mathrm{L^*}$) galaxies, a key question is whether AGN play a role in quenching all the way down to the lowest-mass isolated quiescent galaxies.
    
    Constraining the mechanisms responsible for quenching low-mass galaxies is challenging for a variety of reasons.  The physics underlying AGN feedback remains poorly understood, such that simulations must be tuned to reproduce observed galaxy scaling relations \citep[e.g.,][]{bower2006,somerville2008,gabor2011,genel2014,schaye2015}.  Cosmological simulations are thus limited in their ability to constrain the influence of AGN, particularly in low-mass galaxies which often lie at the limits of resolution. While some zoom-in simulation studies are beginning to include isolated galaxies with $\mstel = 10^7 - 10^9 \ \msun$ \citep[e.g.,][]{graus2019}, there has been minimal focus on AGN and robust observational data to test against remains sparse.
    
    Additionally, identifying AGN in low-mass galaxies remains difficult.  Stellar or gas dynamics are the most secure method for discovering supermassive black holes \citep[for a review see][]{kormendy2004}, but resolution limits largely prohibit the use of this technique beyond the confines of the Local Group.  In more distant systems, it is possible to identify actively accreting systems using broad \ha\ which traces gas orbiting in a deep potential well \citep{greene2007}, and narrow emission line ratios which can distinguish between different sources of ionizing radiation \citep[e.g.,][]{barth2008,reines2013}.  Systematic searches rely on large spectroscopic surveys like SDSS, which are limited by the brightness of host galaxies for targeting, the strength of emission lines for identification, and contamination from the host galaxy which obscures AGN signals. Moreover, while BPT line ratios can be used to identify AGN candidates, without additional follow-up (e.g., X-ray emission, IR colors, etc.) emission cannot be conclusively identified as originating from an AGN.
    
    Despite these challenges, the number of AGN discovered in galaxies with $\mstel < 10^{10} \ \msun$ continues to grow. \citet{reines2013} used the NASA/Sloan Atlas to identify 136 AGN candidates using both broad \ha\ emission and position on the BPT diagram, while \citet{sartori2015} found 336 candidate galaxies that fulfill either mid-IR or BPT criteria. \citet{baldassare2018} recently found an additional sample of 35 AGN in low-mass galaxies using optical variability, many of which would be classified as star-forming from their SDSS fiber-based emission line ratios. 
    
    Recent studies suggest that central black holes may exert significant influence on their low-mass host galaxies. \citet{penny2018} found a set of quiescent galaxies in group environments with both AGN-like BPT line ratios and disturbed gas kinematics, suggesting AGN-driven outflows.  \citet{bradford2018} studied the HI gas masses of isolated low-mass galaxies with BPT-identified AGN and found two populations; composite galaxies with gas content in line with non-AGN galaxies and gas-depleted galaxies with stronger AGN signatures.  They suggest that AGN may be able to affect the cold gas content of some galaxies with low stellar masses.
    
    These studies exist in some tension with other work which suggests that stellar and supernovae feedback, rather than black hole accretion, are the dominant forces which moderate star-formation in low-mass galaxies \citep[e.g.,][]{martinnavarro2018,bower2017}.
    
    In this work, we investigate the processes required to quench isolated galaxies, focusing on the lowest-mass galaxies that are isolated and lack recent star formation. We define galaxies as star-forming or quiescent based on observational properties as measured by the SDSS fiber spectroscopy. We consider galaxies to be quiescent if they have both old stellar populations and sufficiently low levels of star formation. We intentionally differentiate between \textit{quenched} galaxies in which star formation has been fully and permanently disrupted and \textit{quiescent} galaxies which are not currently forming stars but are not necessarily ``dead''.
    
    We discuss the observational proxies and exact definitions used in \autoref{sec:sample} and build a catalog of isolated low-mass galaxies. We present new Keck/ESI long-slit spectroscopy for a subset of these galaxies.  In \autoref{sec:bpt} we measure emission line fluxes for galaxies as a function of radius and place them on the BPT diagram, categorizing individual galaxy spaxels as quiescent, star-forming, AGN-like, or composite. We identify central AGN candidates and classify the radial trends of galaxies on the BPT diagram.  In \autoref{sec:dist}, we measure the perpendicular distance from the star-forming region in [\othree]/\hb\ vs.\ [\ntwo]/\ha\ space, and connect this indicator of the relative strength of non-SF ionizing radiation to host galaxy properties. We summarize our work and its implications in \autoref{sec:summ}.  We assume $\mathrm{H_0 = 70 \ km \ s^{-1} \ Mpc^{-1}}$ throughout this work.

\section{Data}
\label{sec:sample}

\subsection{Galaxy Catalog and Isolation Criterion}

    The parent sample for this work is the NASA/Sloan Atlas\footnote{http://www.nsatlas.org} \citep[NSA; ][]{blanton2011}, a reanalysis of SDSS DR8, optimized for nearby low-luminosity objects. The NSA galaxy catalog includes fluxes, optical line strengths and equivalent widths, and associated errors for all objects derived using the methods of \cite{yan2012} and the SDSS spectrophotometric recalibration of \cite{yan2011}. The catalog also includes stellar masses, estimated with the \texttt{kcorrect} software \citep{blanton2007}, based on SDSS optical fluxes as well as GALEX fluxes where available, and assuming a \cite{chabrier2003} initial mass function.
    
    \begin{figure}
        %\figurenum{text}
        \epsscale{1.1}
        \plotone{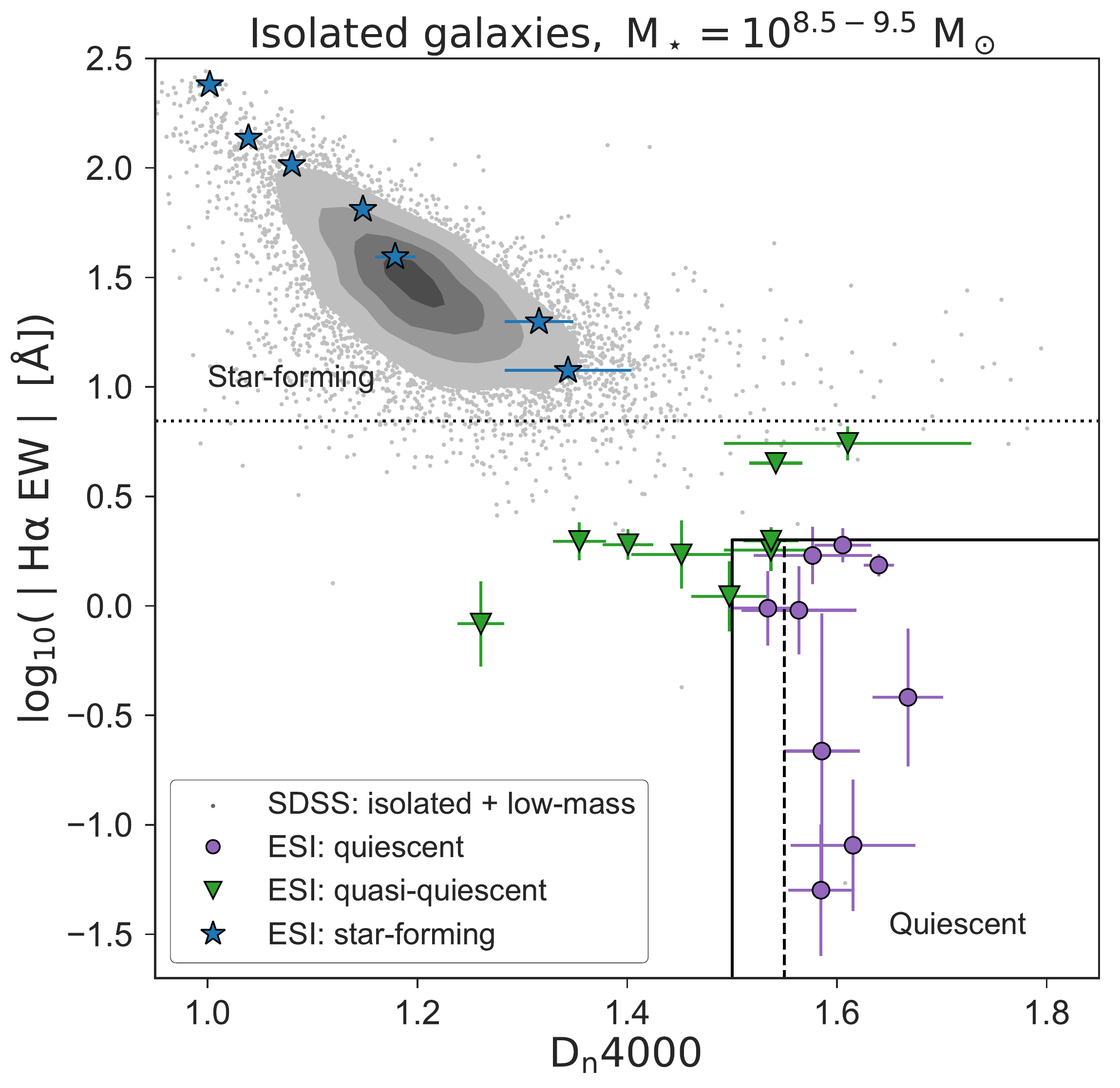}
        \caption{\dnfour \ vs H$\alpha$ equivalent width for isolated ($\mathrm{D}_{\mathrm{host}} > 1.5$ Mpc), low-mass ($\mathrm{\mstel = 10^{8.5} - 10^{9.5} \ \msun}$) galaxy sample (gray points, contours), measured with SDSS. Galaxies for which we have obtained Keck/ESI spectroscopy are shown as colored points and divided into three types based on the strength of their 4000 \AA \ breaks and H$\alpha$ EWs: star-forming (blue stars), quasi-quiescent (green triangles), and quiescent (purple circles). Solid lines indicate the criteria for quiescence (\dnfour\ has a mass dependence, and we show the highest-mass limit with a dashed line). Galaxies are considered to be quasi-quiescent if they lie below the dotted line.}
        \label{fig:dnha}
    \end{figure}
    
    Following \cite{geha2012}, we define galaxies as isolated based on the quantity $D_{\mathrm{host}}$, the projected distance to the nearest ``luminous'' neighbor, where luminous is defined to be $M_{K_s} < -23$ (corresponding to $\mstel = 2.5 \times 10^{10}\ \msun$). Luminous galaxies are considered to be potential host galaxies if they are within 1000 km $\mathrm{s^{-1}}$ in redshift and within a projected comoving distance of 7 Mpc. $D_\mathrm{host}$ is given as the shortest distance to a luminous neighbor within the search parameters. For the small number of galaxies where no luminous neighbor was identified, $D_\mathrm{host}$ was set to 7 Mpc. We consider a galaxy to be isolated if $D_\mathrm{host} > 1.5$ Mpc. 
    
    We focus on galaxies $\pm\ 0.5$ dex around the isolated quenching threshold for a mass range of $\mathrm{\mstel = 10^{8.5} - 10^{9.5} \ \msun}$ \citep{geha2012}. The NSA catalog contains $\mathrm{N = 6850}$ isolated galaxies within this mass range.
    
    Our definition of isolated selects for galaxies that are centrals rather than satellites, but does not discriminate based on large scale environment, unlike other measures (e.g., $3^{\mathrm{rd}}$ or $7^{\mathrm{th}}$ nearest neighbor, aperture counts, etc). Our galaxies all live well beyond the virial radius of their nearest neighbor, but are not necessarily living in true cosmic voids \citep{vanderweygaert2011}.
    
    Within this sample, we further define galaxies as quiescent, quasi-quiescent, or star-forming. We consider a galaxy to be quiescent based on two criteria measured from the SDSS fiber spectra: the $\mathrm{H\alpha}$ equivalent width (EW), which traces the specific star formation rate over the last $\sim 10-20$ Myr, and the $\mathrm{D_n4000}$ index, which is a measure of the light-weighted age of the stellar population and is based on the strength of the 4000 \AA \ break \citep{balogh1999}. Quiescent galaxies are those with $\mathrm{H\alpha \ EW < 2}$ \AA \ and $\mathrm{D_n4000 > 0.6 + 0.1 \cdot \log_{10}(M_*)}$. We base these definitions on the empirical delineations between star-forming and quiescent populations in \citet{geha2012}. We remove 192 galaxies from the sample with poorly measured \dnfour\ ($\mathrm{D_n}4000_{\mathrm{err}} > 0.1$). 
    In the mass range $\mathrm{\mstel = 10^{8.5} - 10^{9.5} \ \msun}$, the majority of galaxies (6470\,/\,6658) are defined as star-forming and just 11 are quiescent, consistent with \citet{geha2012}.   There are an additional 177 galaxies which we define as ``quasi-quiescent'' because they fulfill only one of our criteria for quiescence (large \dnfour \ and intermediate $\mathrm{H\alpha \ EW}$, or low \dnfour \ and low $\mathrm{H\alpha \ EW}$) based on their SDSS fiber-based spectroscopy. These galaxies are potential analogues to the Green Valley observed at higher stellar masses. 

\subsection{Keck/ESI Observations}

    From the above sample, 27 isolated galaxies were observed over four nights (2014 March 6, 2014 March 7, 2015 January 19, and 2016 February 14) with the Echelle Spectrograph and Imager \citep[ESI;][]{sheinis2002} on the Keck II 10m telescope. 
    
    In building our sample, we focused primarily on maximizing the number of quiescent galaxies observed, acquiring ESI observations for 10 of 11 quiescent galaxies, and an additional 10 of 177 quasi-quiescent galaxies which may be in the process of transitioning from star-formation to quiescence. Finally, we also observed 7 galaxies which span the full range of \dnfour\ and \ha\ EWs observed in star-forming, low-mass galaxies in SDSS (\autoref{fig:dnha}, blue stars).
    
    The observations were taken in echellette mode, which provides wavelength coverage over the range 3900-11000 \AA, across 10 echelle orders. The observations were made using the 1.0\arcsec \ x 20\arcsec \ slit, which gives an instrumental resolution of $32 \ \mathrm{km \ s^{-1}}$ (Gaussian $\sigma$) over the full spectrum or $R \approx 10,000$. For each galaxy, the slit was aligned along the major-axis and a minimum of three consecutive 5-minute exposures were obtained. We achieve $\mathrm{\langle SNR\rangle \sim 10-25 \ pix^{-1}}$ in the continuum across the Ca III region.
    
    The ESI data were processed using the \texttt{XIDL} package\footnote{https://www2.keck.hawaii.edu/inst/esi/ESIRedux/index.html} \citep{prochaska2003}, following the ESIRedux cookbook to process the darks, flats, and arc images.  The flat field images were used to trace the edges of the slit across each curved order and to correct gain mismatch across the amplifiers.  From the individual arc images taken on each night, a master CuAr+HgNe+Xe arc image was created.  From this master arc image the wavelength solution was derived for each night of observations.
    
    Once the individual science frames were dark-subtracted and flat-fielded, cosmic rays were identified and removed with the routine \texttt{LACosmic} \citep{vandokkum2001}.  The individual science frames were median-combined and then the initially curved individual orders rectified. 
    
\begin{figure}
        %\figurenum{text}
        \epsscale{1.2}
        \plotone{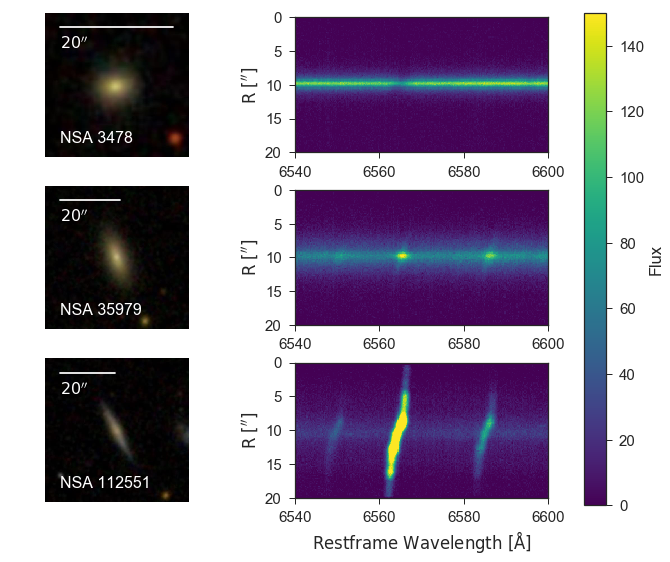}
        \caption{SDSS gri images for 3 galaxies in our sample (left), and corresponding sky-subtracted 2d Keck/ESI spectra for each galaxy, centered on the \ha\ + [\ntwo] region (right). Galaxies in our sample show a variety of emission line profile behaviors, with some dominated by stellar absorption (top), some showing only centrally-concentrated emission (middle), and some with emission line extending across the full slit (bottom). All galaxies shown here have similar stellar masses ($\mstel \sim 10^{9.2 - 9.4} \ \msun$).}
        \label{fig:imgs}
    \end{figure}

    To model the contribution from sky lines and continuum, we performed a bspline fit to a 2\arcsec-wide region on the outer edges of the 2d spectrum. In some cases, emission lines from the galaxy extend across the full slit (e.g., NSA 112551 in \autoref{fig:imgs}). For these galaxies, we use the internal rotation of the galaxy to select uncontaminated regions of sky on the opposite side of the slit. For the few galaxies with negligible rotation, we assume a flat sky continuum and fit to the regions directly surrounding the emission lines in wavelength-space.
    
\subsection{Emission line measurements}

    We are interested in measuring the strength of emission lines in low-mass isolated galaxies such that we can isolate AGN-like line ratios. However, the majority of galaxies in our sample are dominated by the stellar \textit{absorption continuum}. For accurate measurements of \ha\ and \hb\ \textit{emission}, it is necessary to model and remove the effects of stellar absorption. 
    
    For each galaxy in the sample, we fit the stellar continuum from the spectrum extracted within the central 1\arcsec\ (as determined from the median stellar continuum profile at $\lambda = 8400 - 8800$ \AA). After masking the strong optical emission lines, we convolve a set of empirical stellar templates with a velocity profile using pPXF \citep{cappellari2017} and fit to the central galaxy spectrum. 
    
    We have tested pPXF's ability to accurately recover the shape of the Balmer absorption lines by modeling a set of quiescent galaxies with and without emission line region masking. In all cases, the model produced by fitting to the line-masked spectrum is in good agreement with the model fit to the unmasked spectrum. 
    
    We assume that the relative line strength of the absorption features do not vary significantly as a function of radius. Given the stellar continuum derived from the continuum normalized central 1\arcsec, we use the median galaxy light profile to produce a two-dimensional (2d) map of the galaxy continuum.
    
    We subtract the galaxy continuum model from the 2d sky-subtracted spectra to correct for stellar absorption. This produces a 2d emission-line-only spectrum, from which we can extract spatially resolved one-dimensional spectra for each galaxy. The central spectrum is centered on the peak of the galaxy brightness profile and extracted from a 1\arcsec-wide region (corresponding to the average seeing across our nights). Successive spectra are extracted at increasing radii with a width of 1\arcsec.
    
    Given a set of spectra for each galaxy, we follow \citet{reines2013} and model the emission lines of interest as Gaussians. We initially fit the \ha\ + [\ntwo] region, keeping the relative separations fixed at their laboratory wavelengths and the flux ratio of [\ntwo] $\lambda$6583 to [\ntwo] $\lambda$6548 at the intrinsic ratio of 2.96. We then measure the [\stwo] $\lambda\lambda$6718,6732 doublet, \hb and the [\othree] $\lambda\lambda$4959,5007 doublet, while fixing the model to \ha\ + [\ntwo] linewidth and velocity. %\todo{possible additions: errorbars (how to describe?) and choice of fixed line widths across different lines.}
    
    We also test for the presence of a broad \ha\ component by refitting the \ha\ + [\ntwo] region, allowing for an additional broad Gaussian in the \ha\ line with a FWHM of at least 500 $\mathrm{km \ s^{-1}}$. We consider the fit to be improved if the $\chi^2$ increases by 25\%, but do not find any evidence of broad \ha\ for any galaxies within our sample.

%\todo{discussion of broad \ha\ in literature (not necessarily detectable at low-mass?)}

\section{Placing Galaxies on the BPT diagram}
\label{sec:bpt}

    \begin{figure*}
        %\figurenum{text}
        \epsscale{1.15}
        \plotone{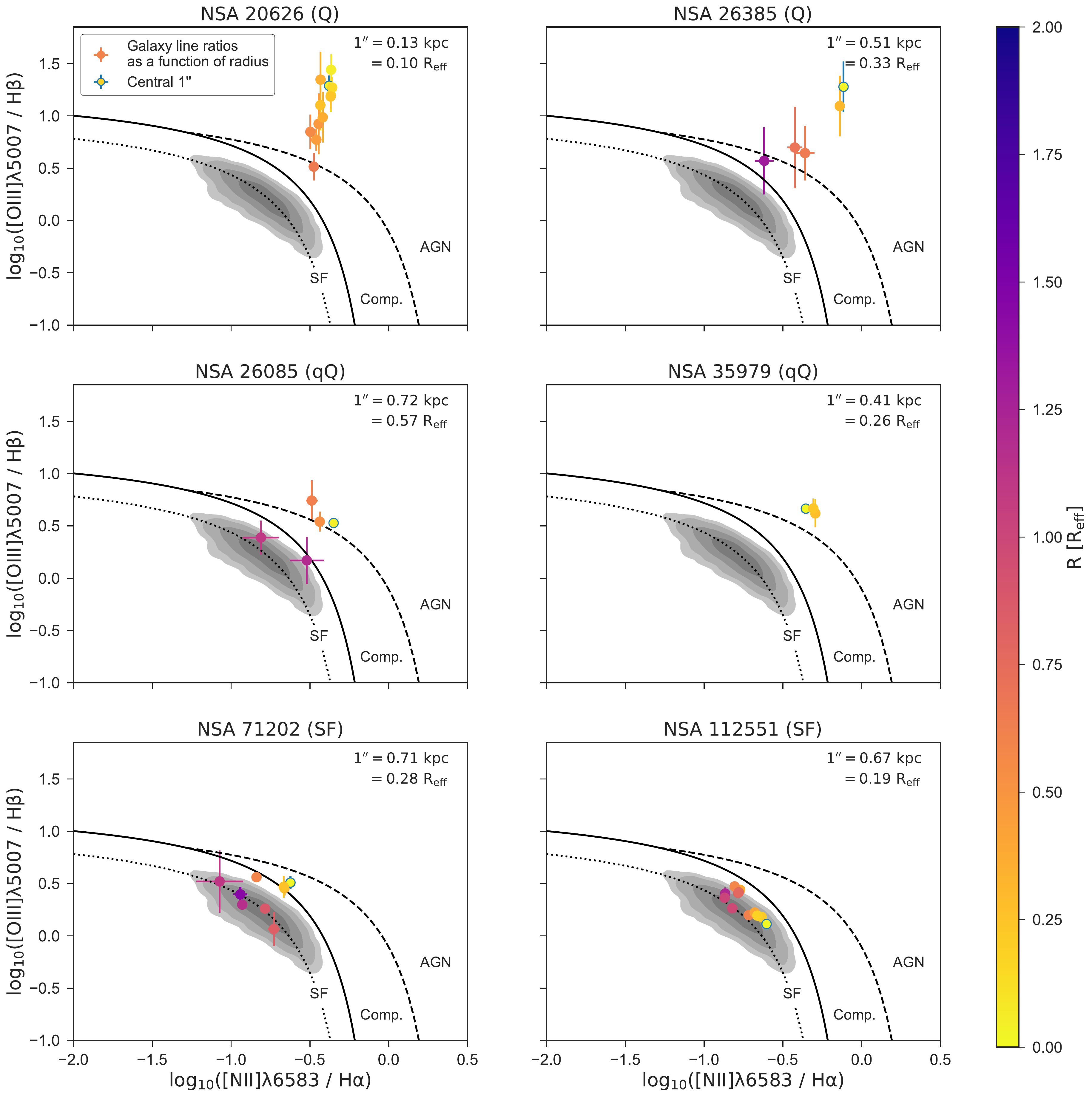}
        \caption{A selection of BPT diagrams for 6 galaxies in our sample which highlight the observed spatial patterns within each galaxy subtype. Each panel shows the locus of low-mass isolated galaxies with well-measured BPT emission lines (\ha\ and \hb\ SNR > 3) from the NASA/Sloan Atlas (gray contours) as well as the demarcation lines between sources of ionizing radiation as given in \citet{kewley2013}, where the dotted line traces the median \htwo\ abundance sequence, the solid line is the empirical delineation between star-formation and AGN \citep{kauffmann2003}, and the dashed line is the maximum starburst line \citep{kewley2001}. We show BPT measurements derived from the ESI spectra as circles, color-coded by distance from the galaxy center. Points are outlined in blue when they lie within the central arcsec of the galaxy. Galaxies fall into three broad categories: dominated by star-formation (bottom row), central AGN with extended SF (top right, middle right), and pure AGN (top left). 16 / 20 galaxies defined as quiescent based on their \dnfour \, and \ha\ EWs show signatures of AGN, along with 2 of 7 star-forming galaxies.}
        \label{fig:bpts}
    \end{figure*}
    
    The BPT diagram \citep{baldwin1981} is a powerful diagnostic for differentiating between the sources of ionizing radiation in a galaxy spectrum \citep{veilleux1987, kewley2001, kewley2006, kauffmann2003}. While there are many variations on the original optical line ratio diagram and many classification schemes therein, we focus on the ``traditional'' BPT, which uses the ratios [\ntwo]/\ha\ vs [\othree]/H$\beta$. 
    
    In this space, star-forming galaxies which are dominated by emission from \htwo\ regions create the sequence marked by a dotted line in each panel of \autoref{fig:bpts} \citep[the star-forming sequence;][]{kewley2013}. Lower metallicities produce higher [\othree]/H$\beta$ ratios and lower [\ntwo]/\ha\ ratios \citep{kewley2006, groves2006, cann2019}. 
    
    We follow the commonly-used classification scheme for distinguishing between sources of ionizing radiation \citep[star formation vs.\ AGN, shocks, or pAGB stars; e.g.,][]{kewley2006, kewley2013, reines2013, baldassare2018}. The dashed line in \autoref{fig:bpts} from \citet{kewley2001} represents the division between a theoretical maximum-starburst model vs.\ emission that requires a harder ionizing spectrum (e.g., AGN activity). The \citet{kauffmann2003} division (solid line in \autoref{fig:bpts}) is an empirical separation between the star-forming sequence and the AGN plume, which extends to larger [\ntwo]/\ha\ and [\othree]/\hb\ ratios. Galaxies below the \citet{kauffmann2003} demarcation are considered to be purely star-forming, while galaxies above the \citet{kewley2001} are traditionally considered to be Seyfert AGN and LINERs. Galaxies lying between these two regions are considered to be composites, whose spectra suggest a mix of ionization from star-formation and harder radiation.
    
    The BPT diagram is primarily used for identifying AGN activity in galaxies, as the harder ionizing radiation from an actively accreting supermassive black hole produces line ratios distinct from those found in a star-forming galaxies. However, there are several other astrophysical process that can produce line ratios similar to those from AGN. In general, BPT line ratios alone are insufficient to rule out other ionization mechanisms, like SNe shocks or heating from pAGB stars \citep{yan2012, yan2018}. 

\subsection{Comparison with SDSS}

    We first compare the SDSS 3\arcsec-fiber flux ratios to those measured from our ESI spectra. We only evaluate the SDSS BPT diagram for a galaxy where the flux measurements have sufficient SNR in the four key lines (\ha, \hb, [\othree], and [\ntwo] SNR > 3). Of the 27 galaxies in our sample, only 12 have reliably measured BPT flux ratios in SDSS. 
    
    For each of the 12 galaxies, we extract a 3\arcsec diameter spectrum from across the ESI slit, centered on the peak of the stellar continuum. The ESI extraction is matched to SDSS in diameter but not area (due to the mismatch of fiber vs. long slit). Still, for all 12 galaxies, we confirm the SDSS BPT classifications within R = 3\arcsec in our ESI data, although we show in \autoref{sec:spatial} centered on the peak of the stellar continuum that this varies with smaller apertures.

\subsection{Spatially-Resolved BPT Diagrams}
\label{sec:spatial}

    The Keck/ESI spectrograph provides up to 20\arcsec\ of spatial information on each galaxy in our sample, with $\sim 1\arcsec$ resolution. In \autoref{fig:bpts}, we present spatially-resolved BPT diagrams for a subset of galaxies in our sample (BPT diagrams for the full sample are included in \autoref{appendix}).
    
    The galaxies in our sample were originally classified as star-forming (7/27), quiescent (10/27), or quasi-quiescent (10/27) based on the SDSS-fiber measurements of their stellar populations (\ha\ EW and \dnfour, as shown in \autoref{fig:dnha}). 
    
    In the 7 galaxies defined as star-forming, the emission lines extend across the full 20\arcsec\ slit (e.g., NSA 112551, bottom row in \autoref{fig:imgs}). All seven galaxies are classified as BPT-SF (emission originating from \htwo\ regions) from the 3\arcsec \ spectra, but two of seven galaxies have composite line ratios within the central 1\arcsec, suggesting contribution to the emission from a non-SF-driven source of hard ionizing radiation. For the remaining five galaxies, the line ratios at all radii are consistent with ionizing radiation originating exclusively from star-formation. 
    
    Though all seven star-forming galaxies are broadly classified as BPT-SF at (almost) all radii, the distribution of the emission line ratios within BPT-SF space as a function of radius varies from galaxy to galaxy. For three of the seven galaxies the line ratios at all radii are loosely clustered, with no clear radial trends. 
    
    In contrast, three galaxies show large radial variations in [\ntwo]/\ha\ (e.g., NSA 112551 in \autoref{fig:bpts}). In two cases, the [\ntwo]/\ha\ ratio is greater at the center of the galaxy and decreases at larger radii, while the third galaxy (NSA 84573, see \autoref{fig:bpt_sf}) has [\ntwo]/\ha\ ratios that increase with radius, while still remaining within the SF region. For all three galaxies, [\othree]/\hb\ is approximately constant as a function of radius. The galaxies not included in \autoref{fig:bpts} are shown in \autoref{appendix}.
    
    Before examining the radial trends for quiescent and quasi-quiescent galaxies, we model the effects of key stellar population parameters on BPT-position, to better disentangle the behavior of underlying galaxy properties from any possible AGN contributions. The quiescent and quasi-quiescent galaxies will be discussed below in \autoref{sec:fsps}.

\subsection{Modelling BPT-behavior with Simple Stellar Populations}
\label{sec:fsps}
    
    \begin{figure}
        %\figurenum{text}
        \plotone{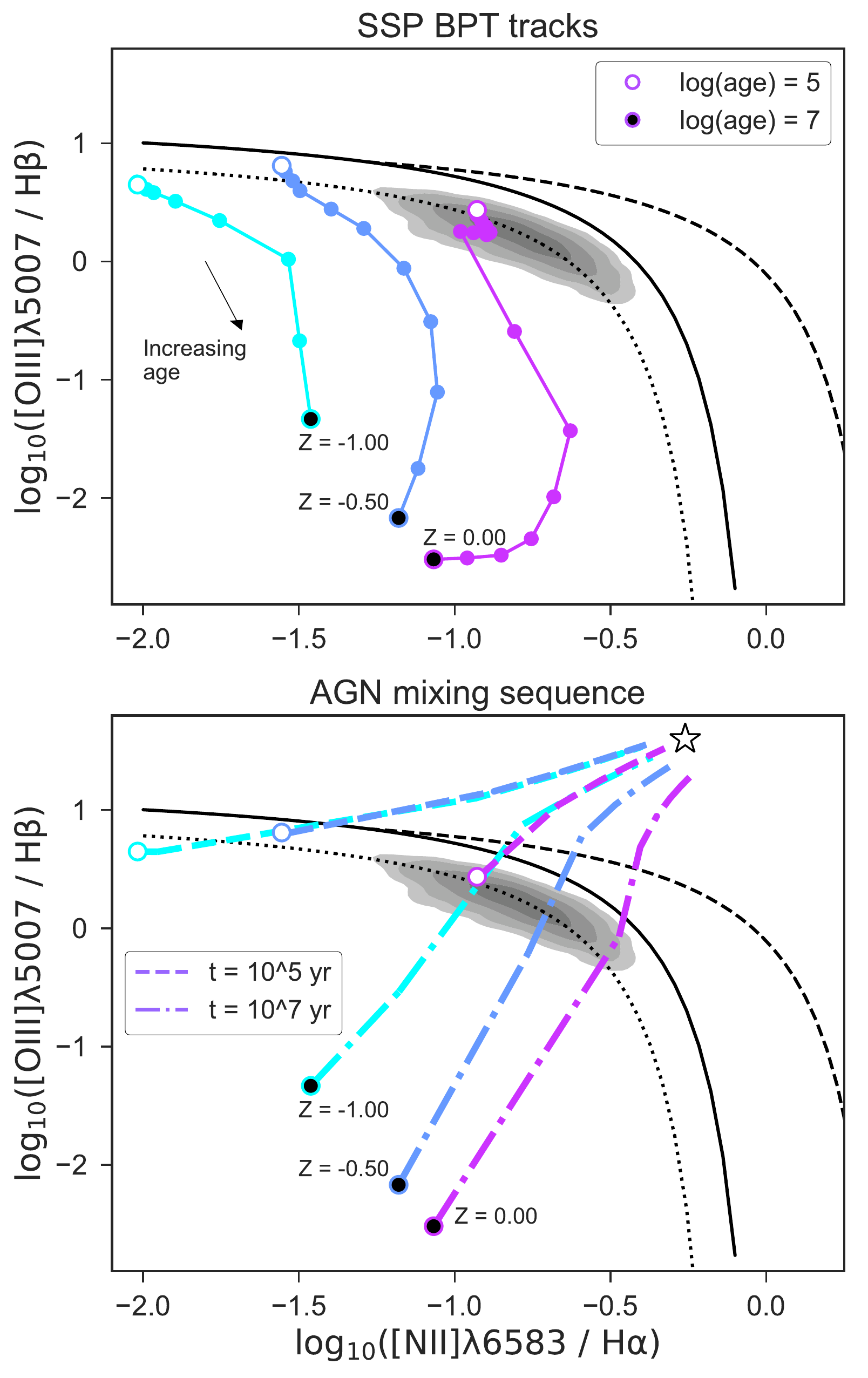}
        \caption{\textit{Upper panel:} The evolution of simple stellar populations in BPT-space as a function of stellar age (solid lines) and metallicity (color). Increasing metallicity moves galaxies to larger [\ntwo]/\ha, while older stellar populations have lower [\othree]/\hb. \\
        \textit{Lower panel:} The effect of adding Seyfert AGN-like emission to a simple stellar population. Dot-dashed lines show the effect of increasing AGN contribution at fixed age and metallicity for the underlying stellar population. The AGN fraction increases from 0 to 90\% towards the upper right. The white star marks our fiducial AGN ratios.}
        \label{fig:fsps}
    \end{figure}
    
    To better understand how galaxies move through BPT-space, we investigate the effects of varying the stellar age, metallicity, and contribution from AGN to the emission line spectra for a set of simple stellar populations. 
    We use \texttt{FSPS} \citep{conroy2009,conroy2010} to generate spectra for simple stellar populations, varying both stellar age and metallicity (solid lines in \autoref{fig:fsps}, upper panel). We assume a \citet{kroupa2001} IMF and set the gas-phase metallicity to match the stellar, with a fixed gas ionization parameter of $\mathrm{\log(U) = -2.0}$. Nebular emission lines are generated for a given SSP as described in \citet{byler2017}.
    
    Actively star-forming galaxies (those with the youngest stellar ages, white-faced circles in \autoref{fig:fsps}) lie along the median \htwo\ abundance sequence. Increasing metallicity moves the spectrum to larger [\ntwo]/\ha\ at nearly constant [\othree]/\hb. 
    
    The distribution of our simple stellar populations through BPT-space suggest that metallicity gradients as a function of radius can explain the horizontal sequences observed in some of the star-forming galaxies. Some cosmological zoom-in simulations predict that potential fluctuations from stellar feedback in low-mass galaxies can drive  negative metallicity gradients \citep[$\mathrm{\Delta\log{Z/Z_\msun} \sim -0.25,}$][]{elbadry2016}, which are echoed by MaNGA observations that have found flat to slightly negative metallicity gradients in low-mass galaxies \citep{belfiore2017}.
    
    We also model the AGN narrow line spectrum as a set of Gaussian emission lines, with the \ha\ and \hb \ fluxes scaled to 10-100\% of the SF fluxes, and [\othree] and [\ntwo] fluxes set such that the ``AGN'' lies in the pure AGN region of the BPT diagram. 
    
    Increasing the flux contribution from the AGN relative to the star-formation produces the curved diagonal tracks shown in \autoref{fig:fsps} (lower panel, dashed and dot-dashed lines). The curvature of these tracks varies with both the metallicity and time elapsed since star-formation, as well as the fundamental line ratios of the AGN. We show how the emission line ratios evolve with increasing AGN fraction for a Seyfert AGN (\autoref{fig:fsps}, bottom panel). While this is  a toy model, it suggests that variation in the contribution from an AGN produces diagonal sequences which are distinct from the horizontal sequences driven by metallicity gradients.
    
    The star-forming galaxies in our sample show little to no signs of AGN activity, with none of the strong radial gradients seen in \autoref{fig:fsps}. The 20 quiescent and quasi-quiescent galaxies in our sample display very different trends. 
    
    Of the 10 quasi-quiescent galaxies (green triangles, \autoref{fig:dnha}), 6/10 have evidence for some on-going star-formation within the slit. However, only one of these 6 galaxies is purely star-forming in BPT-space; the 5 other galaxies show evidence of emission originating from a non-SF-ing source within $\sim 0.5 \ \mathrm{R_{eff}}$ (e.g., NSA 26085, center-left in \autoref{fig:bpts}). In all 4 of the remaining quasi-quiescent galaxies, we only observe emission which is consistent with a non-SF source of ionizing radiation (e.g., NSA 35979, center-right in \autoref{fig:bpts}). 
    
    Finally, our sample includes 10 quiescent galaxies (purple circles in \autoref{fig:dnha}). Of those 10, we find two galaxies (NSA 3478 and 18953) to be fully quiescent systems, with no detected emission at any radii. Of the 8 remaining galaxies for which we can measure the BPT line ratios, only one galaxy (NSA 119887) shows emission consistent with star-formation. Visual inspection of the galaxy in SDSS gri shows a possible off-center \htwo\ region. Given the  small spatial extent and weakness of the \ha\ flux from the  star-forming region relative to central non-SF source, we still consider this galaxy to be pre-dominantly quiescent. 
    
    The 7 other quiescent galaxies exclusively inhabit the composite and AGN-like regions of the BPT diagram, with no evidence of purely star-forming line ratios at any radii. 
    
    In 12 of 15 quiescent and quasi-quiescent galaxies with non-SF line ratios, the AGN-like emission is not contained within the central 1\arcsec. In most cases, AGN-like or composite line ratios are observed out to $R \sim 0.5 \ \mathrm{R_{eff}}$. This is in contrast to the star-forming galaxies with AGN signatures, where galaxies only lie in the BPT-comp./AGN regions in the central-most spaxel. 
    
    This suggests that either AGN in quiescent galaxies have more spatially-extended hard ionizing radiation, or that there is little to no star-formation obscuring the full spatial extent of the harder radiation in these galaxies. The extended narrow line regions may be produced by extended periods of AGN activity or enabled by lower gas densities in quiescent, gas-depleted galaxies.
    
    Of the galaxies which appear to host non-SF-like ionizing radiation, 8\,/\,18 show strong variation in their line ratios as a function of radius. In each of these cases, the galaxy moves diagonally through BPT-space, with both [\ntwo]/\ha\ and [\othree]/\hb\ increasing towards the galaxy center (e.g., NSA 26385, top-right in \autoref{fig:bpts}).  
    
    In the case of NSA 20626 (top-left, \autoref{fig:bpts}), the emission line ratios create a nearly vertical sequence in BPT-space.  Increasing the AGN contribution to the spectrum always changes both the [\othree]/\hb and [\ntwo]/\ha, suggesting that the near-vertical sequence observed in NSA 20626 cannot be explained exclusively by variation in the ratio of star-formation to AGN present in the galaxy.

    To summarize, we have examined the radial behavior of galaxies on the BPT diagram and found that \textbf{the majority of quiescent and quasi-quiescent galaxies have extended AGN-like line ratios which form characteristic diagonal tracks across BPT-space}, which are distinct from the metallicity gradients observed in star-forming galaxies.

\subsection{The Central 1\arcsec}
\label{sec:one_arc}

    \begin{figure}
        %\figurenum{text}
        \epsscale{1.2}
        \plotone{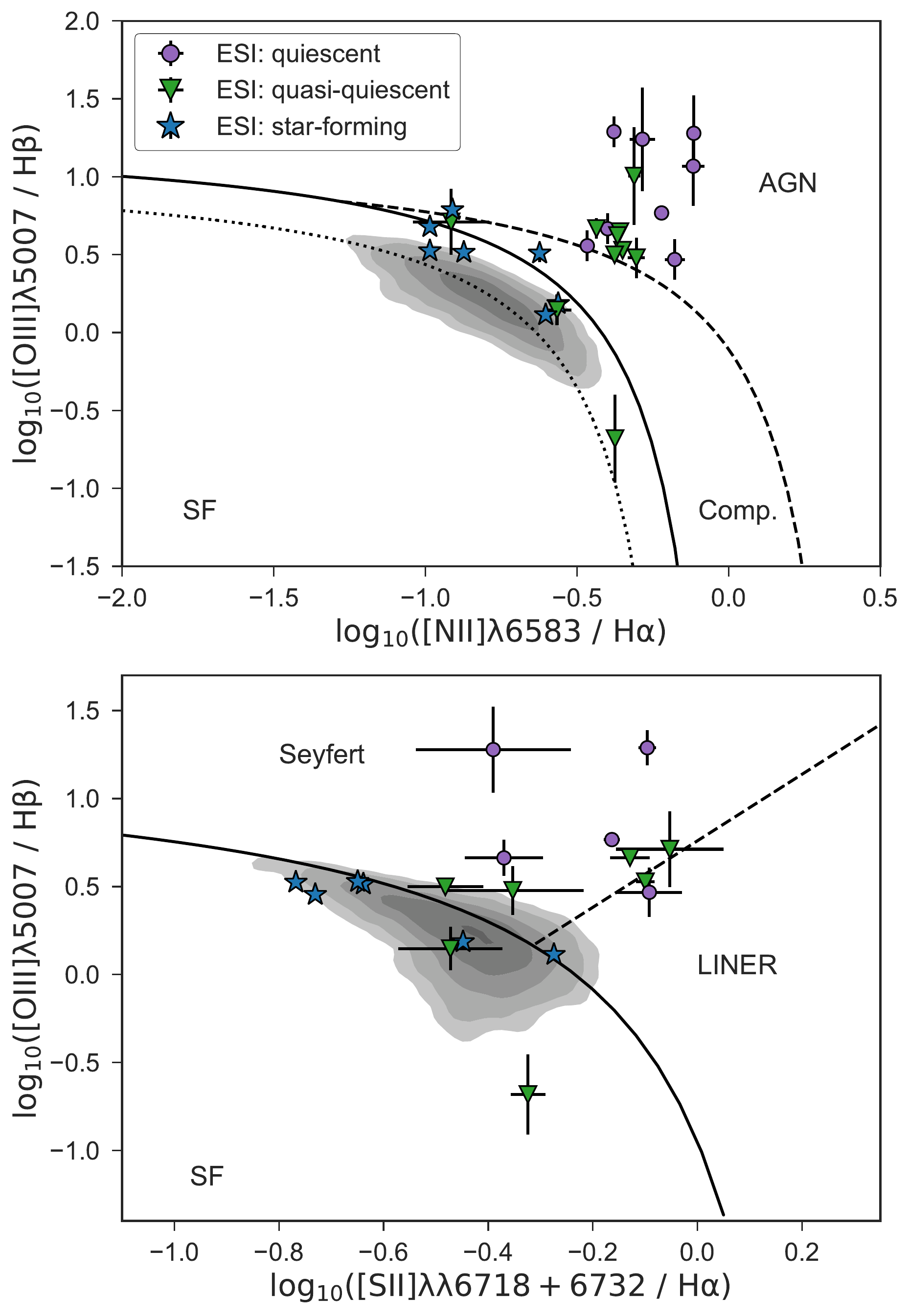}
        \caption{\textit{Upper panel:} The [\ntwo]-BPT diagram for all galaxies in the sample, measured within the central 1\arcsec of the Keck/ESI spectra. Galaxies classified as star-forming from their stellar populations in SDSS (blue stars) lie in the star-forming or composite regions of the BPT diagram. Quiescent and quasi-quiescent galaxies (circles and triangles) lie primarily in the AGN region of the diagram.\\
        \textit{Lower panel:} The [\stwo]-BPT diagram for a subset of galaxies within the sample, once again measured within the central 1\arcsec of the Keck/ESI spectra. The [\stwo]-BPT diagram can be used to distinguish between Seyfert AGN and Low-Ionization Nuclear Emission Regions (LINERs), which may have more ambiguous origins.}
        \label{fig:onearc_bpt}
    \end{figure}

    Our investigation of the spatially-resolved BPT diagram for low-mass galaxies shows that the excitation contribution from non-SF sources can vary significantly as a function of radius, but always peaks within the central arcsecond of each galaxy. A possible explanation for this behavior is that we are observing the increasing influence of AGN activity within each galaxy's narrow line region (NLR).
    
    To better quantify the influence of the harder ionizing radiation present within our sample of galaxies, we focus on only the central 1\arcsec\ of each galaxy, where emission from an AGN, if present, should dominate. Based on the BPT classifications of \citet{kewley2006}, we classify the central 1\arcsec\ as quiescent, AGN-like, or star-forming (\autoref{tab:bpt}, BPT type). The BPT diagram traditionally distinguishes between the ``AGN'' and ``composite'' regions, where composite refers to a mix of star-formation and AGN-driven ionization, but this is a empirical divide, calibrated to high-mass galaxies. Given the low metallicities of the galaxies in our sample, lying anywhere in the AGN or composite region is strongly indicative of a non-SF source of ionizing radiation \citep{groves2006, cann2019}. Going forward, we classify all galaxies lying in either the AGN or composite regions as ``AGN-like''.

    \begin{figure*}
        %\figurenum{text}
        %\epsscale{1.1}
        \plotone{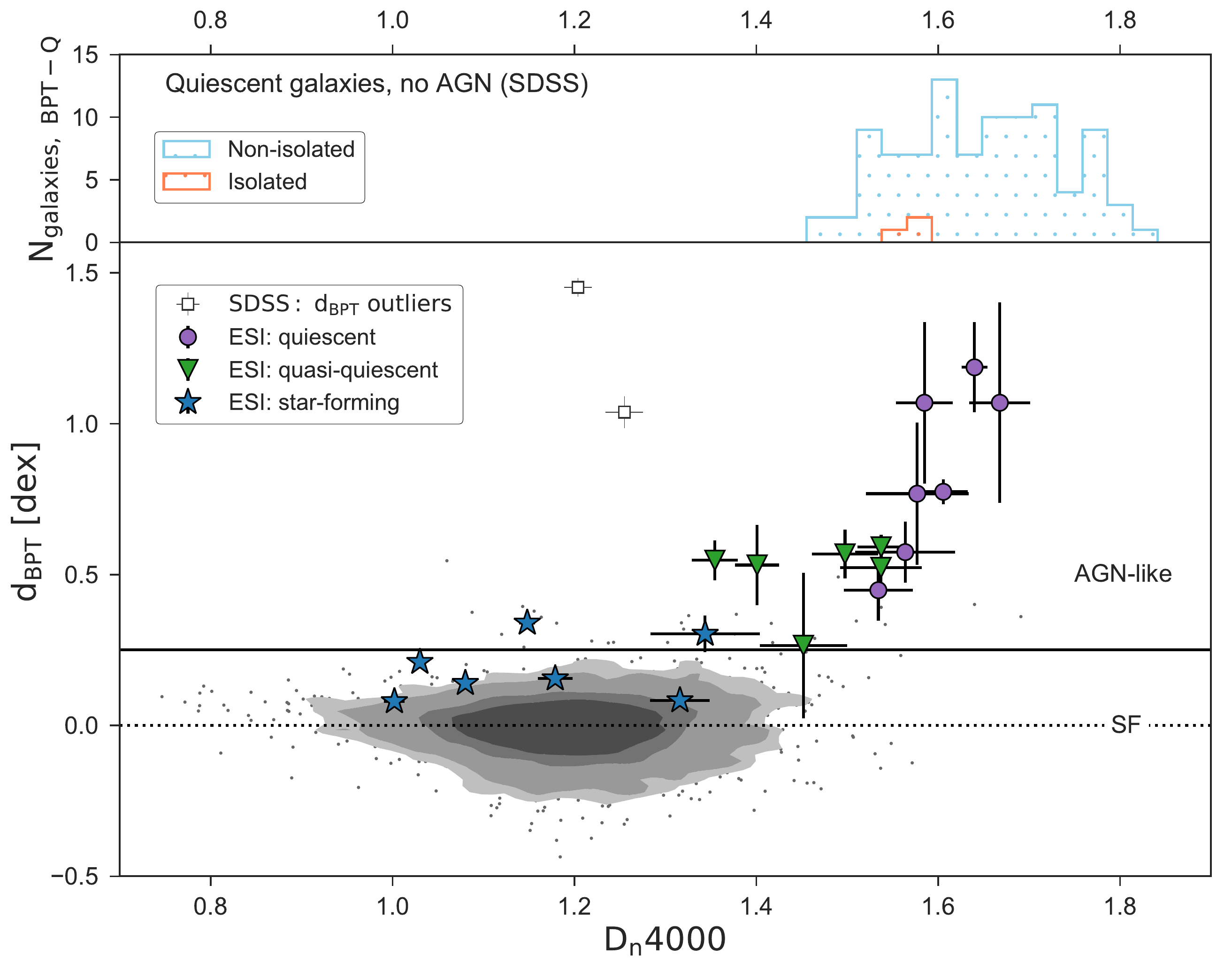}
        \caption{\textit{Lower panel}: $\mathrm{d_{BPT}}$ as a function of \dnfour\ for isolated, low-mass galaxies in SDSS (gray) and galaxies observed with ESI (colors). The \dnfour \, index measures the strength of the Balmer break and is a good proxy for stellar age, with large values of \dnfour\ indicating an evolved stellar population. Large BPT distances indicate stronger ionizing radiation signatures, while $\mathrm{d_{BPT}}$ = 0 indicates a galaxy is on the median star-forming sequence. We find that the BPT-position of the central 1\arcsec\ in a galaxy is tightly correlated with the age of the galaxy. White squares mark two outliers from SDSS with large \dbpt\ measurements but low \dnfour.\\
        \textit{Upper panel}: The number of BPT-quiescent galaxies as a function of \dnfour\ in isolated (orange) and non-isolated (blue) environments. The presence of BPT-Q satellite galaxies in the same mass range indicates that we are not missing a hidden population of BPT-Q isolated galaxies.}
        \label{fig:dbpt}
    \end{figure*}

    \autoref{fig:onearc_bpt} shows the position of each galaxy's central 1\arcsec\ on the [\ntwo] and [\stwo] BPT diagrams (upper and lower panels, respectively). Star-forming galaxies lie near the \htwo\ sequence (dotted line, upper panel). The small number of star-forming galaxies with AGN-like line ratios all lie very close to or within the composite region, suggesting low-metallicity gas \citep{groves2006}, a significant contribution from star-formation in the central 1\arcsec, or both. 
    
    In SDSS, galaxies not classified as star-forming are primarily found in the composite region, where we expect to see galaxies with a mix of star-formation and AGN emission. Within the ESI central 1\arcsec, almost none of the quiescent and quasi-quiescent galaxies lie in this composite region, suggesting that either there is no underlying star-formation at the smallest radii or that we are probing the region where the harder ionizing radiation is strong enough to dominate over any residual star-formation. 
    
    The majority of the quiescent and quasi-quiescent galaxies in our sample lacked sufficient SNR for placement on the BPT diagram with SDSS. In 13 of 15 galaxies, the increased spectral and spatial resolution of the Keck/ESI spectra reveals the presence of faint central emission consistent with AGN-like line ratios.
    
    \textbf{Of the 20 galaxies defined as quiescent or quasi-quiescent based on their stellar populations, we find that 16 show evidence of potentially hosting an actively accreting central black hole}.

\section{BPT Distance and Quiescence}
\label{sec:dist}

    Thus far, we have examined how our galaxies would be defined within the traditional BPT diagram. However, this classification scheme has been calibrated for galaxies more massive than our sample. Low-mass galaxies have weaker AGN emission relative to their high-mass counterparts, and their metal-poor nature can significantly influence their position on the BPT diagram, as metallicity is a significant driver of BPT-position for both star-formation and AGN-dominated spectra, as we have shown in \autoref{sec:fsps}. Due to these complicating factors, we find it useful to parameterize each galaxy's relative position on the BPT diagram. We use \dbpt, the perpendicular distance from the \citet{kewley2013} star-forming sequence (\autoref{fig:onearc_bpt}, dotted line) as measured for the central 1\arcsec\ ESI spectrum of each galaxy. 
    
    We compare \dbpt\ to \dnfour, which measures the strength of the 4000 \AA \ break and serves as a good proxy for the age of stellar populations. We use stellar age as traced by \dnfour\ rather than any proxy for star-formation, as we cannot decouple \ha\ emission originating from \htwo\ regions vs.\ that from other sources. 
    Our ESI spectra lack the required sensitivity in the bluest orders to measure \dnfour, so we rely on the values derived from the 3\arcsec\ SDSS fiber. Simulations predict moderate age gradients in low-mass galaxies \citep[$\Delta \mathrm{t} \sim 1$ Gyr within 1 kpc;][]{elbadry2016}. For quiescent galaxies, where we expect older overall ages, \dnfour\ does not vary significantly on Gyr timescales and should have similar values when measured within 1 vs 3\arcsec. 
    
    In \autoref{fig:dbpt} we find a correlation between \dbpt\ and \dnfour, with the oldest galaxies most removed from the BPT star-forming sequence. AGN found in star-forming galaxies are found significantly lower along the AGN mixing sequence relative to their quiescent galaxy counterparts.
    
    For comparison, the gray contours in \autoref{fig:dbpt} show the full sample of low-mass isolated galaxies, with \dbpt\ values calculated from the SDSS emission line fluxes. The majority of galaxies have $\mathrm{D_n4000} < 1.5$ and scatter about \dbpt $\, = 0$. This is in direct contrast to quiescent and quasi-quiescent galaxies, where we see a clear relation between \dbpt\ and \dnfour. 
    
    Though we attempted to randomly select galaxies from the star-forming population, all 7 galaxies lie above \dbpt = 0 even within the 3\arcsec\ fiber, an unfortunate byproduct of small sample size.  
    
    Two galaxies have SDSS-derived \dnfour\ values indicative of star-formation, but significantly elevated \dbpt\ (white squares in \autoref{fig:dbpt}). Both galaxies have SDSS emission line ratios that place them in the LINER region of the BPT diagram. However, the galaxies lie well outside the range of typical BPT-LINER values.   Given these un-physical values, we plan follow-up spectroscopic observations of these galaxies before further classifying.
    
    If we interpret \dbpt\ as an indicator of position along the AGN mixing sequence, then \textbf{the relative dominance of AGN is correlated with the age of stellar populations in low-mass isolated galaxies. This  suggests that AGN play a significant role in the quenching, that is, the disruption of star formation and the maintenance of low or no SF in low-mass galaxies}. 
    
    Thus far, we have relied on the [\ntwo] BPT diagram to disentangle emission from star-formation vs non-\htwo-region sources.  The most commonly invoked explanation for non-SF-driven emission is an actively accreting black hole, but this is not the only possible source for the observed line ratios. The AGN region of the BPT diagram is known to host two distinct populations: Seyfert AGNs and Low-Ionization Emission Regions (commonly referred to as LINERs when found  in  the  nuclear  regions  of  galaxies). The two populations cannot be distinguished in the composite or near-composite regions of the BPT diagram, though Seyfert galaxies tend to have lower [\ntwo]/\ha\ and higher [\othree]/\hb\ than LINERs. 
    
    Shocked gas from outflows can produce AGN-like line ratios, but it is unclear if such low-mass galaxies are capable of producing sufficiently energetic episodes of star-formation \citep{dashyan2018}. Studies of ionization sources on the BPT diagram \citep[e.g.,][]{yan2012, belfiore2017} have found that heating from pAGB stars can mimic a low-accretion-rate black hole and produce LINER-like emission. However, the majority of our AGN-candidates have central emission line ratios that are strongly characteristic of Seyfert AGN, rather than LINERs. Moreover, in the majority of cases, the \ha\ emission does not directly trace the stellar continuum, suggesting that an old stellar population is not the origin of the ionizing radiation. 
    
    \begin{deluxetable*}{lccccccccc}
        \tablecaption{Isolated low-mass galaxies observed with ESI. \\Galaxy type, Mass, \ha EW, and \dnfour\ are all derived from the SDSS fiber, while BPT type and \dbpt\ are measured from the central 1\arcsec of the ESI spectra. \label{tab:bpt}}
        \tablecolumns{5}
        \tablewidth{0pt}
        \tablehead{
        \colhead{ID} & \colhead{$\alpha$} & \colhead{$\delta$} & \colhead{$cz$} &  \colhead{Galaxy Type} & \colhead{Mass} & \colhead{$\mathrm{H}\alpha$ EW} & \colhead{$\mathrm{D_n4000}$} & \colhead{BPT Type} & \colhead{$\mathrm{d_{BPT}}$} \\
        \colhead{} & \colhead{J2000.0} & \colhead{J2000.0} & \colhead{$\mathrm{km\ s^{-1}}$} & \colhead{} & \colhead{$\msun$} & \colhead{\AA} & \colhead{} & \colhead{} & \colhead{dex}
        }
        \startdata
        NSA 81964 & 15:37:39.6 & 26:15:07.9 & 6654 & SF & 9.10 & 64.7 & 1.14 & AGN & $0.34 \pm 0.02$ \\
        NSA 71202 & 14:23:29.4 & 11:02:40.8 & 10414 & SF & 9.14 & 11.89 & 1.34 & AGN & $0.30 \pm 0.06$ \\
        NSA 76069 & 13:38:42.5 & 14:34:43.3 & 12747 & SF & 9.20 & 240.0 & 1.00 & SF & $0.08 \pm 0.01$ \\
        NSA 84573 & 08:59:35.0 & 26:48:14.3 & 1373 & SF & 9.20 & 39.3 & 1.17 & SF & $0.16 \pm 0.02$ \\
        NSA 112551 & 11:29:27.5 & 24:09:55.1 & 9789 & SF & 9.25 & 19.8 & 1.31 & SF & $0.08 \pm 0.02$ \\
        NSA 77944 & 13:28:04.8 & 06:15:57.7 & 9710 & SF & 9.30 & 136.8 & 1.03 & SF & $0.21 \pm 0.02$ \\
        NSA 66890 & 11:16:09.7 & 06:02:47.0 & 11792 & SF & 9.44 & 103.6 & 1.08 & SF & $0.14 \pm 0.01$ \\
        \hline
        NSA 55500 & 11:06:38.2 & 43:23:44.5 & 6199 & qQ & 8.69 & 1.97 & 1.35 & AGN & $0.55 \pm 0.07$  \\
        NSA 89958 & 12:48:19.0 & 35:12:40.3 & 6836 & qQ & 8.99 & 5.53 & 1.61 & SF & $0.13 \pm 0.11$ \\
        NSA 67565 & 14:02:21.2 & 36:36:06.5 & 6995 & qQ & 9.12 & 1.10 & 1.49 & AGN & $0.57 \pm 0.08$ \\
        NSA 106991 & 10:04:23.3 & 23:13:23.3 & 7971 & qQ & 9.18 & 4.48 & 1.54 & AGN & $0.52 \pm 0.04$ \\
        NSA 91686 & 11:04:39.8 & 36:39:22.6 & 6325 & qQ & 9.25 & 0.82 & 1.26 & AGN & $0.89 \pm 0.33$ \\
        NSA 15814 & 08:13:54.0 & 42:45:19.4 & 9306 & qQ & 9.38 & 6.25 & 1.99 & SF & $0.19 \pm 0.06$ \\
        NSA 61814 & 13:42:38.6 & 46:31:40.3 & 9205 & qQ & 9.39 & 1.71 & 1.45 & AGN & $0.27 \pm 0.24$ \\
        NSA 78759 & 14:12:43.7 & 08:22:16.0 & 8567 & qQ & 9.44 & 1.91 & 1.40 & AGN & $0.53 \pm 0.13$ \\
        NSA 35979 & 08:20:13.9 & 30:25:03.0 & 5885 & qQ & 9.46 & 1.98 & 1.53 & AGN & $0.59 \pm 0.04$ \\
        NSA 26085 & 07:37:29.2 & 30:16:04.2 & 1048 & qQ & 9.47 & 1.79 & 1.53 & AGN & $0.52 \pm 0.05$ \\
        \hline
        NSA 122277 & 15:08:10.8 & 16:05:26.7 & 7327 & Q & 9.01 & 0.97 & 1.53 & AGN & $0.45 \pm 0.10$ \\
        NSA 120659 & 14:31:55.1 & 15:02:26.7 & 8291 & Q & 9.29 & 0.38 & 1.66 & AGN & $1.07 \pm 0.33$ \\
        NSA 3478 & 12:27:34.7 & -01:46:45.7 & 7391 & Q & 9.31 & 0.21 & 1.58 & Q & --- \\
        NSA 51306 & 14:07:27.8 & 46:57:03.3 & 7851 & Q & 9.32 & -0.05 & 1.58 & AGN & $1.07 \pm 0.27$\\
        NSA 6831 & 01:38:42.8 & 00:20:53.0 & 5084 & Q & 9.33 & 1.89 & 1.60 & AGN & $0.77 \pm 0.04$ \\
        NSA 26385 & 08:21:23.6 & 41:42:26.9 & 7375 & Q & 9.37 & 0.98 & 2.01 & AGN & $1.21 \pm 0.24$  \\
        NSA 8762 & 01:31:28.7 & 13:37:52.5 & 9043 & Q & 9.41 & 0.95 & 1.56 & AGN & $0.57 \pm 0.10 $ \\
        NSA 20626 & 16:23:35.0 & 45:44:43.4 & 1844 & Q & 9.41 & 1.53 & 1.63 & AGN & $1.19 \pm 0.15$ \\
        NSA 119887 & 12:30:31.3 & 25:18:34.2 & 6680 & Q & 9.43 & 1.69 & 1.57 & AGN & $0.77 \pm 0.24$ \\
        NSA 18953 & 15:45:16.9 & 02:22:54.2 & 1173 & Q & 9.46 & -0.08 & 1.61 & Q & --- \\
        \enddata
    \end{deluxetable*}

\section{Are We Missing Isolated Quiescent Galaxies?}

    \dbpt\ is a useful metric for parameterizing AGN strength in galaxies. However, truly quiescent galaxies (without any line emission) cannot be placed on the BPT diagram and lack \dbpt\ values. To confirm that we are not missing a population of fully quiescent isolated galaxies, we build a comparison sample of low-mass non-isolated galaxies from the NSA catalog. We define non-isolated galaxies as those with $\mathrm{D_{host}} = 0.4 - 1.0$ Mpc. We add a lower bound to $\mathrm{D_{host}}$ to remove any actively merging or cluster galaxies and set the upper bound to ensure an equal number of non-isolated and isolated galaxies.
    
    For both samples, we identify the sub-population of galaxies that are BPT-quiescent and have quiescent stellar populations. Based on the SDSS fiber measurements, we identify 97 non-isolated galaxies which are fully quiescent and only 3 isolated galaxies (this includes NSA 3478 and 18953, which were observed with ESI but lack emission lines entirely).  The large discrepancy between the two populations, despite the fact that they are drawn from equally sized parent samples in the same mass range confirms that SDSS would have identified any missing truly quiescent population of low-mass isolated galaxies. We show the SDSS-defined truly quiescent population for both isolated and satellite galaxies in the upper panel of \autoref{fig:dbpt}.

\section{Discussion}
\label{sec:discussion}

    We find evidence for AGN-like line ratios in the majority of isolated, low-mass, quiescent galaxies observed in this study, suggesting that AGN play a key role in the quenching of isolated low-mass galaxies.
    
    Despite the growing body of evidence for the importance of AGN in low-mass galaxy evolution, the exact mechanisms by which AGN quench low-mass galaxies remain uncertain. 
    
    We suggest two scenarios for the quenching of low-mass galaxies in isolation, given the prevalence of AGN-like line ratios within the quiescent galaxies. In the first, the quiescent galaxies are distinct from the star-forming population. In a small number of galaxies, the AGN permanently removes the majority of the internal gas reservoir, leaving quiescent galaxies significantly gas-depleted and unable to reignite star-formation. The isolated nature of these galaxies suggests that external cold accretion will not be a significant factor.
    
    A second possibility is that all low-mass isolated galaxies can experience quenching, but that it is not a permanent process. In this scenario, AGN still remove gas from galaxies during particularly energetic bursts, but gas remains in the halo rather than being ejected beyond the virial radius. As an episode of AGN activity ends, the gas rains back onto the galaxy over the course of a dynamical time and star-formation begins again. 
    
    The number of quiescent galaxies would be determined by some combination of the duration of AGN activity and the dynamical time of the galaxy. Simulated galaxies in this mass range have dynamical times of approximately 200-300 Myr \citep{elbadry2016}. Estimates for the duration of periods of AGN activity vary by orders of magnitude \citep{hickox2014,schawinski2015}, but suggest a timescale of $10^7$ years or fewer, suggesting that the dynamical timescale should drive the length of each quenching period.

    If we were to assume that every galaxy has experienced a single quenching episode since z = 1, lasting approximately a dynamical time, we would expect a quiescent fraction of $\sim 5\%$. This rough estimate is significantly larger than the volume-corrected observed quiescent fraction of $\sim 0.3\%$ \citep{geha2012} for isolated, low-mass galaxies in the NSA catalog. This suggests that the actual periods of quiescence are much briefer, perhaps due to incomplete removal of gas from the galaxy. Additionally, in this scenario, the quasi-quiescent galaxies should be a mix of galaxies moving to and from quiescence. Star-formation histories for these galaxies should distinguish between galaxies just beginning or ending a quiescent phase.
    
    Alternatively, some form of internal feedback unrelated to AGN could have removed or heated the gas in the quiescent galaxies, disrupting star-formation and reducing the gas density in the inner regions of each galaxy. In turn, this would allow radiation from the AGN to extend much further out, producing the observed spatially extended and elevated AGN-like line ratios. However, any scenario that does not invoke AGN in the quenching process must explain why the majority of quiescent galaxies are observed to have AGN-like line ratios.

\section{Summary}
\label{sec:summ}

    In this work, we investigate the connection between AGN-like line ratios and quiescence in low-mass, isolated galaxies. We have acquired high-resolution, spatially-resolved spectroscopy for 27 isolated low-mass galaxies, and measured key optical emission lines as a function of radius. We are able to place 25 of 27 galaxies on the Baldwin Phillips Terlevich (BPT) diagram in a spatially-resolved manner. Within the central 1\arcsec\ of each galaxy, we measure the galaxy's distance from the star-forming sequence in BPT-space (\dbpt). We use this parameter as a measure of AGN activity within a galaxy. Our results are summarized as follows:
    
    \begin{enumerate}
        \item We qualitatively distinguish a variety of trends in the radial BPT diagram for galaxies in our sample. We find a significant number of quiescent galaxies have extended emission which forms distinct tracks in BPT-space, moving from the star-forming or composite regions at $\mathrm{R \sim R_{eff}}$ to the AGN region in the galaxy center (\autoref{fig:bpts}, as well as \autoref{appendix}).
        \item We identify central non-SF ionizing radiation in 16 of 20 quiescent and quasi-quiescent galaxies (\autoref{fig:onearc_bpt}). 
        \item There is a strong trend between \dnfour\ (tracing the average age of stellar populations) and \dbpt, with the oldest galaxies showing the strongest signatures of AGN (\autoref{fig:dbpt}). 
    \end{enumerate}
    
    While we cannot confirm that AGN are actively quenching the galaxies in our sample, the ubiquity of AGN-like line ratios in low-mass quiescent galaxies and the tight correlation we have found between the presence of elevated AGN-like line ratios and the age of stellar populations in our sample adds to the growing body of evidence that AGN play a crucial role in self-quenching galaxies at all mass scales. Careful follow-up with both radio and X-ray facilities will be crucial for confirming the presence of AGN within these systems.
    
    Additionally, constraining the ages of the quiescent and quasi-quiescent galaxies in our sample will help build up a coherent picture of the quenching process at play. In this work, we have used the 4000-\AA\ break to trace the mean light-weighted age of stellar populations in galaxies. \dnfour\ is an effective method for identifying recent ($\sim 1-3$ Gyr) bursts of star formation, but becomes insensitive to older stellar ages (e.g., 7 vs 13 Gyr), particularly at low metallicities \citep{kauffmann2003_dnfour}. The measurement of spatially-resolved star formation histories for our sample of low-mass isolated galaxies will provide significant insight into the quenching process at work. Future work will explore the spatially-resolved SFHs as well as the kinematics of both the gas and stars as derived from our ESI observations in order to constrain the nature of AGN feedback within low-mass, isolated galaxies. \\

\acknowledgements

CMD thanks Jenny Greene, Vivienne Baldassare, and Michael Tremmel for their helpful comments on this work. MG acknowledges support from the National Science Foundation under AST Grant Number 1517422 and the Howard Hughes Medical Institute (HHMI)Professors Program. AW was supported by a Caltech-Carnegie Fellowship, in part through the Moore Center for Theoretical Cosmology and Physics at Caltech, and by NASA, through ATP grant 80NSSC18K1097 and HST GO-14734 and AR-15057 from STScI. KEB acknowledges support from an NSF graduate research fellowship.

\bibliographystyle{yahapj}

\begin{thebibliography}{}
\providecommand\natexlab[1]{#1}
\providecommand\JournalTitle[1]{#1}

\bibitem[{{Aihara} {et~al.}(2011){Aihara}, {Allende Prieto}, {An}, {Anderson},
  {Aubourg}, {Balbinot}, {Beers}, {Berlind}, {Bickerton}, {Bizyaev}, {Blanton},
  {Bochanski}, {Bolton}, {Bovy}, {Brandt}, {Brinkmann}, {Brown}, {Brownstein},
  {Busca}, {Campbell}, {Carr}, {Chen}, {Chiappini}, {Comparat}, {Connolly},
  {Cortes}, {Croft}, {Cuesta}, {da Costa}, {Davenport}, {Dawson}, {Dhital},
  {Ealet}, {Ebelke}, {Edmondson}, {Eisenstein}, {Escoffier}, {Esposito},
  {Evans}, {Fan}, {Femen{\'{\i}}a Castell{\'a}}, {Font-Ribera}, {Frinchaboy},
  {Ge}, {Gillespie}, {Gilmore}, {Gonz{\'a}lez Hern{\'a}ndez}, {Gott}, {Gould},
  {Grebel}, {Gunn}, {Hamilton}, {Harding}, {Harris}, {Hawley}, {Hearty}, {Ho},
  {Hogg}, {Holtzman}, {Honscheid}, {Inada}, {Ivans}, {Jiang}, {Johnson},
  {Jordan}, {Jordan}, {Kazin}, {Kirkby}, {Klaene}, {Knapp}, {Kneib},
  {Kochanek}, {Koesterke}, {Kollmeier}, {Kron}, {Lampeitl}, {Lang}, {Le Goff},
  {Lee}, {Lin}, {Long}, {Loomis}, {Lucatello}, {Lundgren}, {Lupton}, {Ma},
  {MacDonald}, {Mahadevan}, {Maia}, {Makler}, {Malanushenko}, {Malanushenko},
  {Mandelbaum}, {Maraston}, {Margala}, {Masters}, {McBride}, {McGehee},
  {McGreer}, {M{\'e}nard}, {Miralda-Escud{\'e}}, {Morrison}, {Mullally},
  {Muna}, {Munn}, {Murayama}, {Myers}, {Naugle}, {Neto}, {Nguyen}, {Nichol},
  {O'Connell}, {Ogando}, {Olmstead}, {Oravetz}, {Padmanabhan},
  {Palanque-Delabrouille}, {Pan}, {Pandey}, {P{\^a}ris}, {Percival},
  {Petitjean}, {Pfaffenberger}, {Pforr}, {Phleps}, {Pichon}, {Pieri}, {Prada},
  {Price-Whelan}, {Raddick}, {Ramos}, {Reyl{\'e}}, {Rich}, {Richards}, {Rix},
  {Robin}, {Rocha-Pinto}, {Rockosi}, {Roe}, {Rollinde}, {Ross}, {Ross},
  {Rossetto}, {S{\'a}nchez}, {Sayres}, {Schlegel}, {Schlesinger}, {Schmidt},
  {Schneider}, {Sheldon}, {Shu}, {Simmerer}, {Simmons}, {Sivarani}, {Snedden},
  {Sobeck}, {Steinmetz}, {Strauss}, {Szalay}, {Tanaka}, {Thakar}, {Thomas},
  {Tinker}, {Tofflemire}, {Tojeiro}, {Tremonti}, {Vandenberg}, {Vargas
  Maga{\~n}a}, {Verde}, {Vogt}, {Wake}, {Wang}, {Weaver}, {Weinberg}, {White},
  {White}, {Yanny}, {Yasuda}, {Yeche}, \& {Zehavi}}]{aihara2011}
{Aihara}, H., {Allende Prieto}, C., {An}, D., {et~al.} 2011,
  \href{http://dx.doi.org/10.1088/0067-0049/193/2/29}{\JournalTitle{\apjs},
  193, 29}

\bibitem[{{Baldassare} {et~al.}(2018){Baldassare}, {Geha}, \&
  {Greene}}]{baldassare2018}
{Baldassare}, V.~F., {Geha}, M., \& {Greene}, J. 2018, \JournalTitle{ArXiv
  e-prints}, arXiv:1808.09578

\bibitem[{{Baldwin} {et~al.}(1981){Baldwin}, {Phillips}, \&
  {Terlevich}}]{baldwin1981}
{Baldwin}, J.~A., {Phillips}, M.~M., \& {Terlevich}, R. 1981,
  \href{http://dx.doi.org/10.1086/130766}{\JournalTitle{Publications of the
  Astronomical Society of the Pacific}, 93, 5}

\bibitem[{{Balogh} {et~al.}(1999){Balogh}, {Morris}, {Yee}, {Carlberg}, \&
  {Ellingson}}]{balogh1999}
{Balogh}, M.~L., {Morris}, S.~L., {Yee}, H.~K.~C., {Carlberg}, R.~G., \&
  {Ellingson}, E. 1999,
  \href{http://dx.doi.org/10.1086/308056}{\JournalTitle{\apj}, 527, 54}

\bibitem[{{Barth} {et~al.}(2008){Barth}, {Greene}, \& {Ho}}]{barth2008}
{Barth}, A.~J., {Greene}, J.~E., \& {Ho}, L.~C. 2008,
  \href{http://dx.doi.org/10.1088/0004-6256/136/3/1179}{\JournalTitle{\aj},
  136, 1179}

\bibitem[{{Belfiore} {et~al.}(2017){Belfiore}, {Maiolino}, {Tremonti},
  {S{\'a}nchez}, {Bundy}, {Bershady}, {Westfall}, {Lin}, {Drory}, {Boquien},
  {Thomas}, \& {Brinkmann}}]{belfiore2017}
{Belfiore}, F., {Maiolino}, R., {Tremonti}, C., {et~al.} 2017,
  \href{http://dx.doi.org/10.1093/mnras/stx789}{\JournalTitle{\mnras}, 469,
  151}

\bibitem[{{Blanton} {et~al.}(2011){Blanton}, {Kazin}, {Muna}, {Weaver}, \&
  {Price-Whelan}}]{blanton2011}
{Blanton}, M.~R., {Kazin}, E., {Muna}, D., {Weaver}, B.~A., \& {Price-Whelan},
  A. 2011,
  \href{http://dx.doi.org/10.1088/0004-6256/142/1/31}{\JournalTitle{\aj}, 142,
  31}

\bibitem[{{Blanton} \& {Roweis}(2007)}]{blanton2007}
{Blanton}, M.~R., \& {Roweis}, S. 2007,
  \href{http://dx.doi.org/10.1086/510127}{\JournalTitle{\aj}, 133, 734}

\bibitem[{{Bower} {et~al.}(2006){Bower}, {Benson}, {Malbon}, {Helly}, {Frenk},
  {Baugh}, {Cole}, \& {Lacey}}]{bower2006}
{Bower}, R.~G., {Benson}, A.~J., {Malbon}, R., {et~al.} 2006,
  \href{http://dx.doi.org/10.1111/j.1365-2966.2006.10519.x}{\JournalTitle{\mnras},
  370, 645}

\bibitem[{{Bower} {et~al.}(2017){Bower}, {Schaye}, {Frenk}, {Theuns},
  {Schaller}, {Crain}, \& {McAlpine}}]{bower2017}
{Bower}, R.~G., {Schaye}, J., {Frenk}, C.~S., {et~al.} 2017,
  \href{http://dx.doi.org/10.1093/mnras/stw2735}{\JournalTitle{\mnras}, 465,
  32}

\bibitem[{{Bradford} {et~al.}(2018){Bradford}, {Geha}, {Greene}, {Reines}, \&
  {Dickey}}]{bradford2018}
{Bradford}, J.~D., {Geha}, M.~C., {Greene}, J.~E., {Reines}, A.~E., \&
  {Dickey}, C.~M. 2018,
  \href{http://dx.doi.org/10.3847/1538-4357/aac88d}{\JournalTitle{\apj}, 861,
  50}

\bibitem[{{Byler} {et~al.}(2017){Byler}, {Dalcanton}, {Conroy}, \&
  {Johnson}}]{byler2017}
{Byler}, N., {Dalcanton}, J.~J., {Conroy}, C., \& {Johnson}, B.~D. 2017,
  \href{http://dx.doi.org/10.3847/1538-4357/aa6c66}{\JournalTitle{\apj}, 840,
  44}

\bibitem[{{Cann} {et~al.}(2019){Cann}, {Satyapal}, {Abel}, {Blecha},
  {Mushotzky}, {Reynolds}, \& {Secrest}}]{cann2019}
{Cann}, J.~M., {Satyapal}, S., {Abel}, N.~P., {et~al.} 2019,
  \href{http://dx.doi.org/10.3847/2041-8213/aaf88d}{\JournalTitle{\apj}, 870,
  L2}

\bibitem[{{Cappellari}(2017)}]{cappellari2017}
{Cappellari}, M. 2017,
  \href{http://dx.doi.org/10.1093/mnras/stw3020}{\JournalTitle{\mnras}, 466,
  798}

\bibitem[{{Chabrier}(2003)}]{chabrier2003}
{Chabrier}, G. 2003,
  \href{http://dx.doi.org/10.1086/376392}{\JournalTitle{\pasp}, 115, 763}

\bibitem[{{Choi} {et~al.}(2015){Choi}, {Ostriker}, {Naab}, {Oser}, \&
  {Moster}}]{choi2015}
{Choi}, E., {Ostriker}, J.~P., {Naab}, T., {Oser}, L., \& {Moster}, B.~P. 2015,
  \href{http://dx.doi.org/10.1093/mnras/stv575}{\JournalTitle{\mnras}, 449,
  4105}

\bibitem[{{Conroy} \& {Gunn}(2010)}]{conroy2010}
{Conroy}, C., \& {Gunn}, J.~E. 2010,
  \href{http://dx.doi.org/10.1088/0004-637X/712/2/833}{\JournalTitle{\apj},
  712, 833}

\bibitem[{{Conroy} {et~al.}(2009){Conroy}, {Gunn}, \& {White}}]{conroy2009}
{Conroy}, C., {Gunn}, J.~E., \& {White}, M. 2009,
  \href{http://dx.doi.org/10.1088/0004-637X/699/1/486}{\JournalTitle{\apj},
  699, 486}

\bibitem[{{Croton} {et~al.}(2006){Croton}, {Springel}, {White}, {De Lucia},
  {Frenk}, {Gao}, {Jenkins}, {Kauffmann}, {Navarro}, \& {Yoshida}}]{croton2006}
{Croton}, D.~J., {Springel}, V., {White}, S. D.~M., {et~al.} 2006,
  \href{http://dx.doi.org/10.1111/j.1365-2966.2005.09675.x}{\JournalTitle{\mnras},
  365, 11}

\bibitem[{{Dashyan} {et~al.}(2018){Dashyan}, {Silk}, {Mamon}, {Dubois}, \&
  {Hartwig}}]{dashyan2018}
{Dashyan}, G., {Silk}, J., {Mamon}, G.~A., {Dubois}, Y., \& {Hartwig}, T. 2018,
  \href{http://dx.doi.org/10.1093/mnras/stx2716}{\JournalTitle{\mnras}, 473,
  5698}

\bibitem[{{Dekel} \& {Silk}(1986)}]{dekel1986}
{Dekel}, A., \& {Silk}, J. 1986,
  \href{http://dx.doi.org/10.1086/164050}{\JournalTitle{\apj}, 303, 39}

\bibitem[{{El-Badry} {et~al.}(2016){El-Badry}, {Wetzel}, {Geha}, {Hopkins},
  {Kere{\v{s}}}, {Chan}, \& {Faucher-Gigu{\`e}re}}]{elbadry2016}
{El-Badry}, K., {Wetzel}, A., {Geha}, M., {et~al.} 2016,
  \href{http://dx.doi.org/10.3847/0004-637X/820/2/131}{\JournalTitle{\apj},
  820, 131}

\bibitem[{{Gabor} {et~al.}(2011){Gabor}, {Dav{\'e}}, {Oppenheimer}, \&
  {Finlator}}]{gabor2011}
{Gabor}, J.~M., {Dav{\'e}}, R., {Oppenheimer}, B.~D., \& {Finlator}, K. 2011,
  \href{http://dx.doi.org/10.1111/j.1365-2966.2011.19430.x}{\JournalTitle{\mnras},
  417, 2676}

\bibitem[{{Geha} {et~al.}(2012){Geha}, {Blanton}, {Yan}, \&
  {Tinker}}]{geha2012}
{Geha}, M., {Blanton}, M.~R., {Yan}, R., \& {Tinker}, J.~L. 2012,
  \href{http://dx.doi.org/10.1088/0004-637X/757/1/85}{\JournalTitle{\apj}, 757,
  85}

\bibitem[{{Genel} {et~al.}(2014){Genel}, {Vogelsberger}, {Springel}, {Sijacki},
  {Nelson}, {Snyder}, {Rodriguez-Gomez}, {Torrey}, \& {Hernquist}}]{genel2014}
{Genel}, S., {Vogelsberger}, M., {Springel}, V., {et~al.} 2014,
  \href{http://dx.doi.org/10.1093/mnras/stu1654}{\JournalTitle{\mnras}, 445,
  175}

\bibitem[{{Graus} {et~al.}(2019){Graus}, {Bullock}, {Fitts}, {Cooper},
  {Boylan-Kolchin}, {Weisz}, {Wetzel}, {Feldmann}, {Faucher-Gigu{\`e}re},
  {Quataert}, {Hopkins}, \& {Keres}}]{graus2019}
{Graus}, A.~S., {Bullock}, J.~S., {Fitts}, A., {et~al.} 2019,
  \JournalTitle{arXiv e-prints}, arXiv:1901.05487

\bibitem[{{Greene} \& {Ho}(2007)}]{greene2007}
{Greene}, J.~E., \& {Ho}, L.~C. 2007,
  \href{http://dx.doi.org/10.1086/522082}{\JournalTitle{\apj}, 670, 92}

\bibitem[{{Groves} {et~al.}(2006){Groves}, {Heckman}, \&
  {Kauffmann}}]{groves2006}
{Groves}, B.~A., {Heckman}, T.~M., \& {Kauffmann}, G. 2006,
  \href{http://dx.doi.org/10.1111/j.1365-2966.2006.10812.x}{\JournalTitle{\mnras},
  371, 1559}

\bibitem[{{Hickox} {et~al.}(2014){Hickox}, {Mullaney}, {Alexander}, {Chen},
  {Civano}, {Goulding}, \& {Hainline}}]{hickox2014}
{Hickox}, R.~C., {Mullaney}, J.~R., {Alexander}, D.~M., {et~al.} 2014,
  \href{http://dx.doi.org/10.1088/0004-637X/782/1/9}{\JournalTitle{\apj}, 782,
  9}

\bibitem[{{Kauffmann} {et~al.}(2003{\natexlab{a}}){Kauffmann}, {Heckman},
  {White}, {Charlot}, {Tremonti}, {Brinchmann}, {Bruzual}, {Peng}, {Seibert},
  {Bernardi}, {Blanton}, {Brinkmann}, {Castander}, {Cs{\'a}bai}, {Fukugita},
  {Ivezic}, {Munn}, {Nichol}, {Padmanabhan}, {Thakar}, {Weinberg}, \&
  {York}}]{kauffmann2003_dnfour}
{Kauffmann}, G., {Heckman}, T.~M., {White}, S. D.~M., {et~al.}
  2003{\natexlab{a}},
  \href{http://dx.doi.org/10.1046/j.1365-8711.2003.06291.x}{\JournalTitle{\mnras},
  341, 33}

\bibitem[{{Kauffmann} {et~al.}(2003{\natexlab{b}}){Kauffmann}, {Heckman},
  {Tremonti}, {Brinchmann}, {Charlot}, {White}, {Ridgway}, {Brinkmann},
  {Fukugita}, {Hall}, {Ivezi{\'c}}, {Richards}, \& {Schneider}}]{kauffmann2003}
{Kauffmann}, G., {Heckman}, T.~M., {Tremonti}, C., {et~al.} 2003{\natexlab{b}},
  \href{http://dx.doi.org/10.1111/j.1365-2966.2003.07154.x}{\JournalTitle{\mnras},
  346, 1055}

\bibitem[{{Kewley} {et~al.}(2001){Kewley}, {Dopita}, {Sutherland}, {Heisler},
  \& {Trevena}}]{kewley2001}
{Kewley}, L.~J., {Dopita}, M.~A., {Sutherland}, R.~S., {Heisler}, C.~A., \&
  {Trevena}, J. 2001,
  \href{http://dx.doi.org/10.1086/321545}{\JournalTitle{\apj}, 556, 121}

\bibitem[{{Kewley} {et~al.}(2006){Kewley}, {Groves}, {Kauffmann}, \&
  {Heckman}}]{kewley2006}
{Kewley}, L.~J., {Groves}, B., {Kauffmann}, G., \& {Heckman}, T. 2006,
  \href{http://dx.doi.org/10.1111/j.1365-2966.2006.10859.x}{\JournalTitle{\mnras},
  372, 961}

\bibitem[{{Kewley} {et~al.}(2013){Kewley}, {Maier}, {Yabe}, {Ohta}, {Akiyama},
  {Dopita}, \& {Yuan}}]{kewley2013}
{Kewley}, L.~J., {Maier}, C., {Yabe}, K., {et~al.} 2013,
  \href{http://dx.doi.org/10.1088/2041-8205/774/1/L10}{\JournalTitle{\apj},
  774, L10}

\bibitem[{{Kormendy} \& {Kennicutt}(2004)}]{kormendy2004}
{Kormendy}, J., \& {Kennicutt}, Robert~C., J. 2004,
  \href{http://dx.doi.org/10.1146/annurev.astro.42.053102.134024}{\JournalTitle{Annual
  Review of Astronomy and Astrophysics}, 42, 603}

\bibitem[{{Kroupa}(2001)}]{kroupa2001}
{Kroupa}, P. 2001,
  \href{http://dx.doi.org/10.1046/j.1365-8711.2001.04022.x}{\JournalTitle{\mnras},
  322, 231}

\bibitem[{{Mart{\'\i}n-Navarro} \& {Mezcua}(2018)}]{martinnavarro2018}
{Mart{\'\i}n-Navarro}, I., \& {Mezcua}, M. 2018,
  \href{http://dx.doi.org/10.3847/2041-8213/aab103}{\JournalTitle{\apj}, 855,
  L20}

\bibitem[{{Papovich} {et~al.}(2018){Papovich}, {Kawinwanichakij}, {Quadri},
  {Glazebrook}, {Labb{\'e}}, {Tran}, {Forrest}, {Kacprzak}, {Spitler},
  {Straatman}, \& {Tomczak}}]{papovich2018}
{Papovich}, C., {Kawinwanichakij}, L., {Quadri}, R.~F., {et~al.} 2018,
  \href{http://dx.doi.org/10.3847/1538-4357/aaa766}{\JournalTitle{\apj}, 854,
  30}

\bibitem[{{Pasquali} {et~al.}(2010){Pasquali}, {Gallazzi}, {Fontanot}, {van den
  Bosch}, {De Lucia}, {Mo}, \& {Yang}}]{pasquali2010}
{Pasquali}, A., {Gallazzi}, A., {Fontanot}, F., {et~al.} 2010,
  \href{http://dx.doi.org/10.1111/j.1365-2966.2010.17074.x}{\JournalTitle{\mnras},
  407, 937}

\bibitem[{{Peng} {et~al.}(2010){Peng}, {Lilly}, {Kova{\v{c}}}, {Bolzonella},
  {Pozzetti}, {Renzini}, {Zamorani}, {Ilbert}, {Knobel}, {Iovino}, {Maier},
  {Cucciati}, {Tasca}, {Carollo}, {Silverman}, {Kampczyk}, {de Ravel},
  {Sanders}, {Scoville}, {Contini}, {Mainieri}, {Scodeggio}, {Kneib}, {Le
  F{\`e}vre}, {Bardelli}, {Bongiorno}, {Caputi}, {Coppa}, {de la Torre},
  {Franzetti}, {Garilli}, {Lamareille}, {Le Borgne}, {Le Brun}, {Mignoli},
  {Perez Montero}, {Pello}, {Ricciardelli}, {Tanaka}, {Tresse}, {Vergani},
  {Welikala}, {Zucca}, {Oesch}, {Abbas}, {Barnes}, {Bordoloi}, {Bottini},
  {Cappi}, {Cassata}, {Cimatti}, {Fumana}, {Hasinger}, {Koekemoer},
  {Leauthaud}, {Maccagni}, {Marinoni}, {McCracken}, {Memeo}, {Meneux}, {Nair},
  {Porciani}, {Presotto}, \& {Scaramella}}]{peng2010}
{Peng}, Y.-j., {Lilly}, S.~J., {Kova{\v{c}}}, K., {et~al.} 2010,
  \href{http://dx.doi.org/10.1088/0004-637X/721/1/193}{\JournalTitle{\apj},
  721, 193}

\bibitem[{{Penny} {et~al.}(2018){Penny}, {Masters}, {Smethurst}, {Nichol},
  {Krawczyk}, {Bizyaev}, {Greene}, {Liu}, {Marinelli}, {Rembold}, {Riffel},
  {Ilha}, {Wylezalek}, {Andrews}, {Bundy}, {Drory}, {Oravetz}, \&
  {Pan}}]{penny2018}
{Penny}, S.~J., {Masters}, K.~L., {Smethurst}, R., {et~al.} 2018,
  \href{http://dx.doi.org/10.1093/mnras/sty202}{\JournalTitle{\mnras}, 476,
  979}

\bibitem[{{Prochaska} {et~al.}(2003){Prochaska}, {Gawiser}, {Wolfe}, {Cooke},
  \& {Gelino}}]{prochaska2003}
{Prochaska}, J.~X., {Gawiser}, E., {Wolfe}, A.~M., {Cooke}, J., \& {Gelino}, D.
  2003, \href{http://dx.doi.org/10.1086/375839}{\JournalTitle{The Astrophysical
  Journal Supplement Series}, 147, 227}

\bibitem[{{Reines} {et~al.}(2013){Reines}, {Greene}, \& {Geha}}]{reines2013}
{Reines}, A.~E., {Greene}, J.~E., \& {Geha}, M. 2013,
  \href{http://dx.doi.org/10.1088/0004-637X/775/2/116}{\JournalTitle{\apj},
  775, 116}

\bibitem[{{Sartori} {et~al.}(2015){Sartori}, {Schawinski}, {Treister},
  {Trakhtenbrot}, {Koss}, {Shirazi}, \& {Oh}}]{sartori2015}
{Sartori}, L.~F., {Schawinski}, K., {Treister}, E., {et~al.} 2015,
  \href{http://dx.doi.org/10.1093/mnras/stv2238}{\JournalTitle{\mnras}, 454,
  3722}

\bibitem[{{Schawinski} {et~al.}(2015){Schawinski}, {Koss}, {Berney}, \&
  {Sartori}}]{schawinski2015}
{Schawinski}, K., {Koss}, M., {Berney}, S., \& {Sartori}, L.~F. 2015,
  \href{http://dx.doi.org/10.1093/mnras/stv1136}{\JournalTitle{\mnras}, 451,
  2517}

\bibitem[{{Schaye} {et~al.}(2015){Schaye}, {Crain}, {Bower}, {Furlong},
  {Schaller}, {Theuns}, {Dalla Vecchia}, {Frenk}, {McCarthy}, {Helly},
  {Jenkins}, {Rosas-Guevara}, {White}, {Baes}, {Booth}, {Camps}, {Navarro},
  {Qu}, {Rahmati}, {Sawala}, {Thomas}, \& {Trayford}}]{schaye2015}
{Schaye}, J., {Crain}, R.~A., {Bower}, R.~G., {et~al.} 2015,
  \href{http://dx.doi.org/10.1093/mnras/stu2058}{\JournalTitle{\mnras}, 446,
  521}

\bibitem[{{Sheinis} {et~al.}(2002){Sheinis}, {Bolte}, {Epps}, {Kibrick},
  {Miller}, {Radovan}, {Bigelow}, \& {Sutin}}]{sheinis2002}
{Sheinis}, A.~I., {Bolte}, M., {Epps}, H.~W., {et~al.} 2002,
  \href{http://dx.doi.org/10.1086/341706}{\JournalTitle{\pasp}, 114, 851}

\bibitem[{{Smith} {et~al.}(2012){Smith}, {Lucey}, {Price}, {Hudson}, \&
  {Phillipps}}]{smith2012}
{Smith}, R.~J., {Lucey}, J.~R., {Price}, J., {Hudson}, M.~J., \& {Phillipps},
  S. 2012,
  \href{http://dx.doi.org/10.1111/j.1365-2966.2011.19956.x}{\JournalTitle{\mnras},
  419, 3167}

\bibitem[{{Somerville} \& {Dav{\'e}}(2015)}]{somerville2015}
{Somerville}, R.~S., \& {Dav{\'e}}, R. 2015,
  \href{http://dx.doi.org/10.1146/annurev-astro-082812-140951}{\JournalTitle{Annual
  Review of Astronomy and Astrophysics}, 53, 51}

\bibitem[{{Somerville} {et~al.}(2008){Somerville}, {Hopkins}, {Cox},
  {Robertson}, \& {Hernquist}}]{somerville2008}
{Somerville}, R.~S., {Hopkins}, P.~F., {Cox}, T.~J., {Robertson}, B.~E., \&
  {Hernquist}, L. 2008,
  \href{http://dx.doi.org/10.1111/j.1365-2966.2008.13805.x}{\JournalTitle{\mnras},
  391, 481}

\bibitem[{{Su} {et~al.}(2018){Su}, {Hopkins}, {Hayward}, {Ma},
  {Faucher-Gigu{\`e}re}, {Kere{\v{s}}}, {Orr}, \& {Robles}}]{su2018}
{Su}, K.-Y., {Hopkins}, P.~F., {Hayward}, C.~C., {et~al.} 2018,
  \JournalTitle{arXiv e-prints}, arXiv:1809.09120

\bibitem[{{van de Weygaert} \& {Platen}(2011)}]{vanderweygaert2011}
{van de Weygaert}, R., \& {Platen}, E. 2011,
  \href{http://dx.doi.org/10.1142/S2010194511000092}{in International Journal
  of Modern Physics Conference Series, Vol.~1}, 41

\bibitem[{{van Dokkum}(2001)}]{vandokkum2001}
{van Dokkum}, P.~G. 2001,
  \href{http://dx.doi.org/10.1086/323894}{\JournalTitle{Publications of the
  Astronomical Society of the Pacific}, 113, 1420}

\bibitem[{{Veilleux} \& {Osterbrock}(1987)}]{veilleux1987}
{Veilleux}, S., \& {Osterbrock}, D.~E. 1987,
  \href{http://dx.doi.org/10.1086/191166}{\JournalTitle{The Astrophysical
  Journal Supplement Series}, 63, 295}

\bibitem[{{Wetzel} {et~al.}(2013){Wetzel}, {Tinker}, {Conroy}, \& {van den
  Bosch}}]{wetzel2013}
{Wetzel}, A.~R., {Tinker}, J.~L., {Conroy}, C., \& {van den Bosch}, F.~C. 2013,
  \href{http://dx.doi.org/10.1093/mnras/stt469}{\JournalTitle{\mnras}, 432,
  336}

\bibitem[{{White} \& {Frenk}(1991)}]{white1991}
{White}, S. D.~M., \& {Frenk}, C.~S. 1991,
  \href{http://dx.doi.org/10.1086/170483}{\JournalTitle{\apj}, 379, 52}

\bibitem[{{White} \& {Rees}(1978)}]{white1978}
{White}, S.~D.~M., \& {Rees}, M.~J. 1978,
  \href{http://dx.doi.org/10.1093/mnras/183.3.341}{\JournalTitle{\mnras}, 183,
  341}

\bibitem[{{Yan}(2011)}]{yan2011}
{Yan}, R. 2011,
  \href{http://dx.doi.org/10.1088/0004-6256/142/5/153}{\JournalTitle{\aj}, 142,
  153}

\bibitem[{{Yan}(2018)}]{yan2018}
---. 2018,
  \href{http://dx.doi.org/10.1093/mnras/sty2143}{\JournalTitle{\mnras}, 481,
  476}

\bibitem[{{Yan} \& {Blanton}(2012)}]{yan2012}
{Yan}, R., \& {Blanton}, M.~R. 2012,
  \href{http://dx.doi.org/10.1088/0004-637X/747/1/61}{\JournalTitle{\apj}, 747,
  61}

\end{thebibliography}

\appendix
\section{Spatially-Resolved BPT Diagrams}
\label{appendix}

\begin{figure}[H]
%\figurenum{text}
\epsscale{1.15}
\plotone{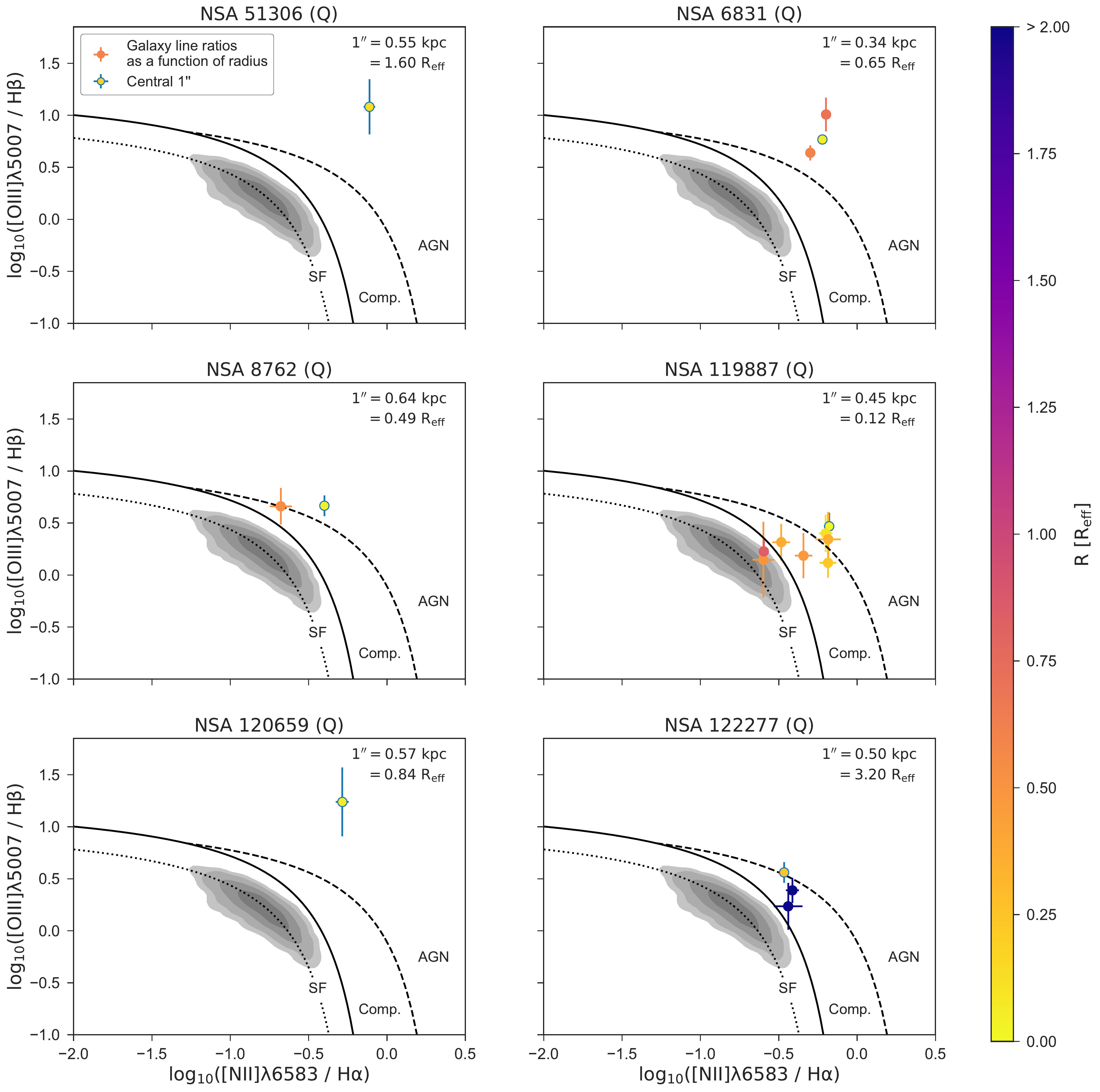}
\caption{BPT diagrams for quiescent galaxies in our sample. Each panel shows the locus of low-mass isolated galaxies with well-measured BPT emission lines (\ha\ and \hb\ SNR > 3) from the NASA/Sloan Atlas (gray contours) as well as the demarcation lines between sources of ionizing radiation as given in \citet{kewley2013}, where the dotted line traces the median [\htwo] abundance sequence, the solid line is the empirical delineation between star-formation and AGN \citep{kauffmann2003}, and the dashed line is the maximum starburst line \citep{kewley2001}. We show BPT measurements derived from the ESI spectra as circles, color-coded by distance from the galaxy center.}
\label{fig:bpt_Q}
\end{figure}

\begin{figure}[H]
%\figurenum{text}
\epsscale{1.15}
\plotone{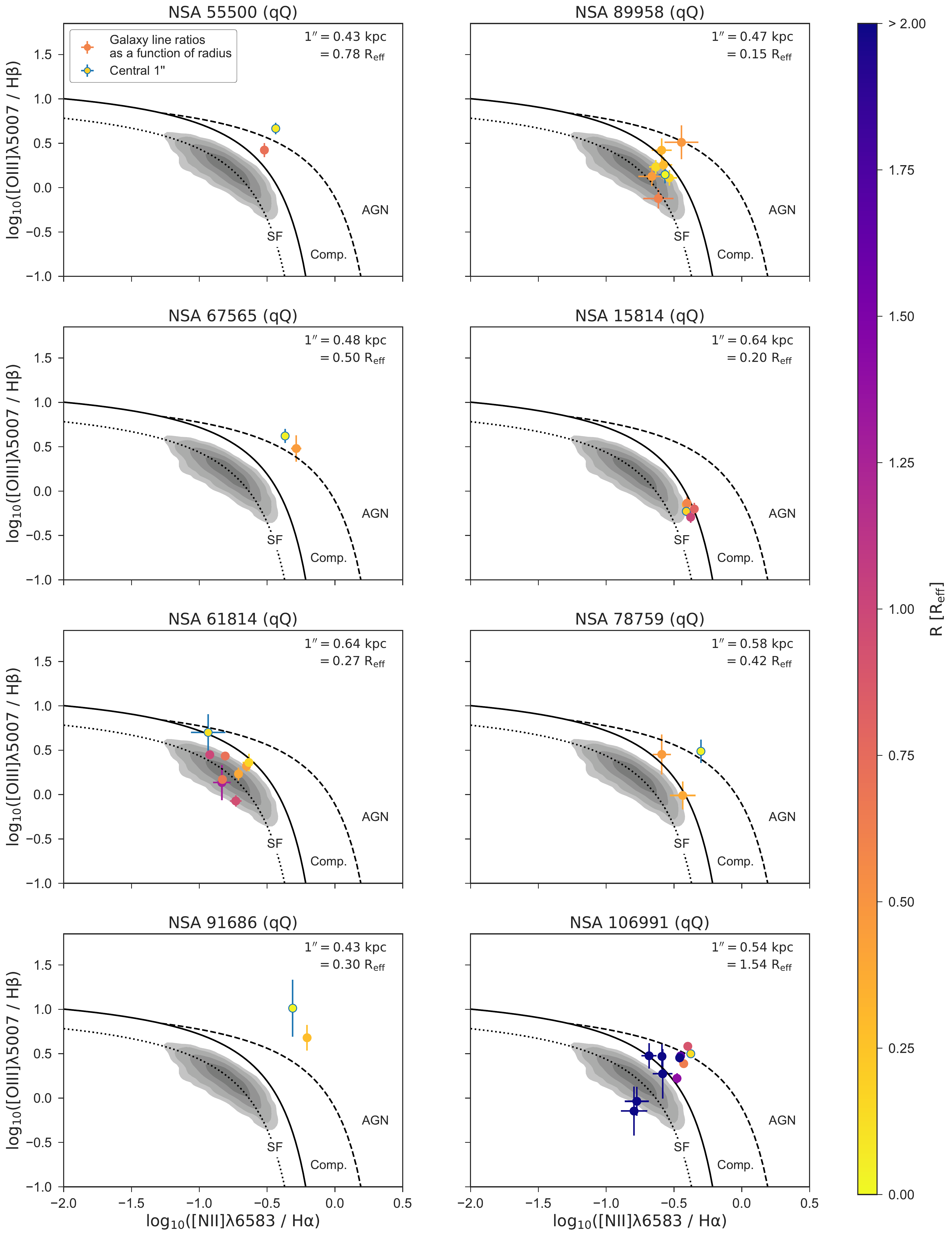}
\caption{Same as \autoref{fig:bpt_Q}, for quasi-quiescent galaxies.}
\label{fig:bpt_qQ}
\end{figure}

\begin{figure}[H]
%\figurenum{text}
\epsscale{1.15}
\plotone{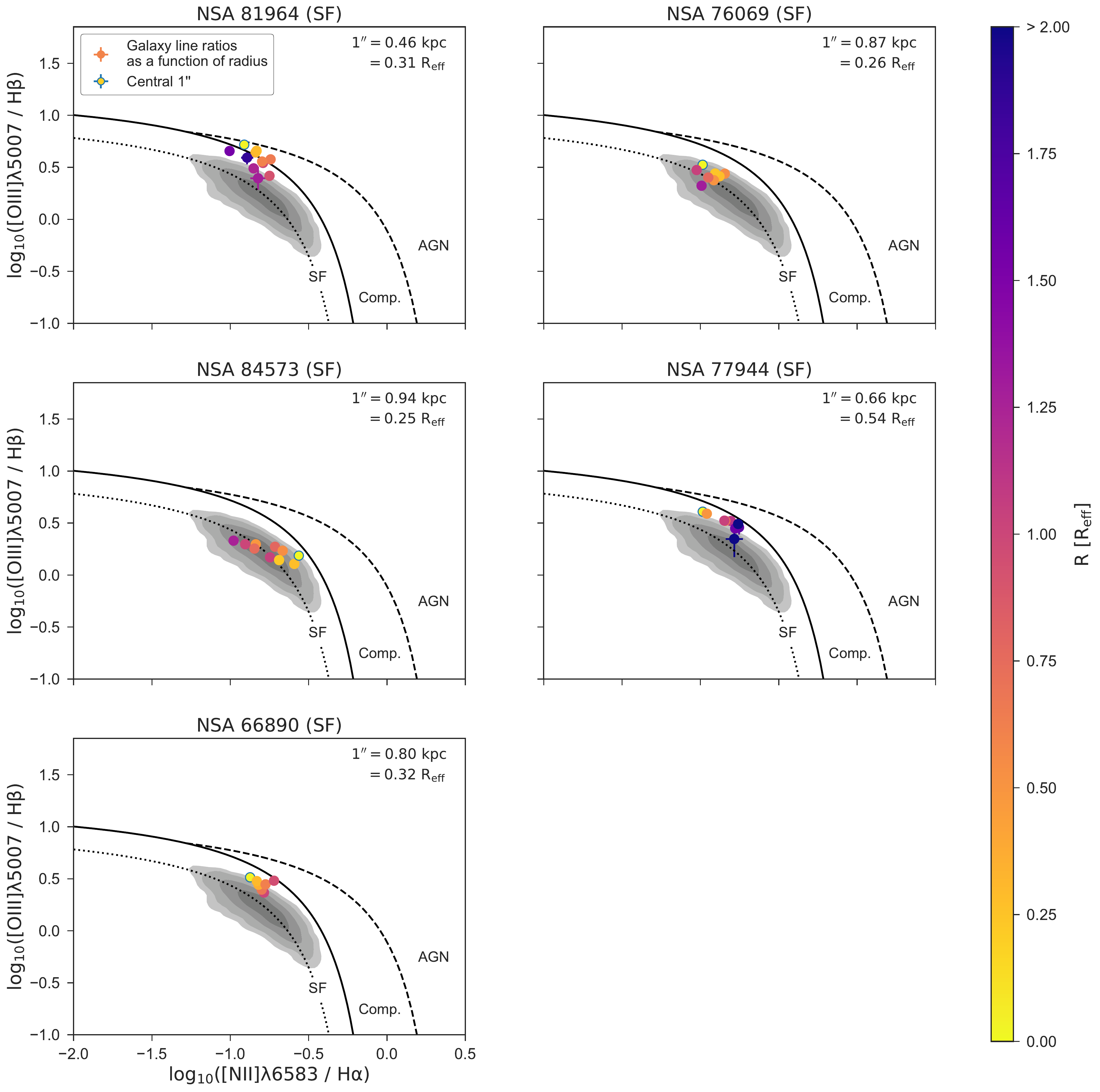}
\caption{Same as \autoref{fig:bpt_Q}, for star-forming galaxies.}
\label{fig:bpt_sf}
\end{figure}

\end{document}